%% file: main.tex
\documentclass[a4paper,11pt]{article}
\pdfoutput=1 

\input{header/header.tex}

\abstract{Many high-energy-physics (HEP) simulations for the LHC rely on Monte Carlo using importance sampling by means of the VEGAS algorithm. However, complex high-precision calculations have become a challenge for the standard toolbox, as this approach suffers from poor performance in complex cases.
As a result, there has been keen interest in HEP for modern machine learning to power adaptive sampling. While previous studies have shown the potential of normalizing-flow-powered neural importance sampling (NIS) over VEGAS, there remains a gap in accessible tools tailored for non-experts. In response, we introduce \zunis, a fully automated NIS library designed to bridge this divide, while at the same time providing the infrastructure to customise the algorithm for dealing with challenging tasks.
After a general introduction on NIS, we first show how to extend the original formulation of NIS to reuse samples over multiple gradient steps while guaranteeing a stable training, yielding a significant improvement for slow functions. Next, we introduce the structure of the library, which can be used by non-experts with minimal effort and is extensivly documented, which is crucial to become a mature tool for the wider HEP public. We present systematic benchmark results on both toy and physics examples, and stress the benefit of providing different survey strategies, which allows higher performance in challenging cases. We show that \zunis shows high performance on a range of problems with limited fine-tuning. }

\author{Nicolas Deutschmann$^{1}$}
\author{Niklas G\"otz$^{2,3}$}
\affiliation{$^1$ IBM Research GmbH, 8803 Rüschlikon, Switzerland}
\affiliation{$^2$Goethe University Frankfurt, Department of Physics, Institute for Theoretical Physics, 60438 Frankfurt, Germany}
\affiliation{$^3$Frankfurt Institute for Advanced Studies, Ruth-Moufang-Strasse 1, 60438
Frankfurt am Main, Germany}

\emailAdd{goetz@itp.uni-frankfurt.de}

\begin{document}
\maketitle
\flushbottom
\section{Introduction}
\label{sec:intro}
\input{introduction/introduction.tex}

\section{Background}
\label{sec:background}
\subsection{Importance sampling as an optimization problem}
\label{sec:importance_sampling}
\input{introduction/importance_sampling.tex}

\subsection{Normalizing flows and coupling cells}
\label{sec:normalizing_flows}
\input{introduction/normalizing_flows}

\subsection{Neural importance sampling}
\label{sec:neural_importance_sampling}
\input{introduction/neural_importance_sampling.tex}

\section{Concepts and algorithms}
\label{sec:framework}
In this section we describe the original contributions of this paper.
The major conceptual innovation we provide in \zunis is a more flexible and data-efficient way of training normalizing flows in the context of importance sampling.
This relies on a more rigorous formulation of the connection between the theoretical expression of ideal loss functions in terms of integrals and their practical realisations as random estimators than in previous literature, combined with an adaptive strategy to switch between these realistions. We describe this improvement in \cref{sec:training}.
We also give a high-level overview of the organisation of the \zunis library, which implements this new training procedure.
\subsection{Efficient training for importance sampling}
\label{sec:training}
\input{framework/training.tex}

\subsection{The \zunis library}
\input{framework/framework.tex}

\section{Results}
\label{sec:performance}
\input{performance/perf.tex}
\subsection{Low-dimensional examples}
\label{sec:qualitative}
\input{performance/qualitative.tex}
\subsection{Systematic benchmarks}
\label{sec:benchmarks}
\input{performance/benchmarks.tex}
\subsection{\textsc{MadGraph} cross section integrals}
\input{performance/madgraph_iclr.tex}
\label{sec:madgraph}

\FloatBarrier
\section{Conclusion}
\input{conclusion/conclusion.tex}
\section{Acknowledgements}
\input{conclusion/acknowledgements.tex}
\section{Reproducibility Statement}
\input{conclusion/reproducibility.tex}

\bibliography{all}
\newpage
\appendix

\section{Event generation with the veto algorithm}
\label{app:unweighting}
\input{appendix/unweighting.tex}

\section{Backward Sampling Algorithm}
\label{app:backward_sampling_alg}
\begin{algorithm}[H]
\SetAlgoLined
\KwData{Parametric PDF $p(x,\theta)$}
\KwResult{Trained PDF $p(x,\theta)$}
 \For{M steps}{
    Sample a batch $x_1,\dots,x_{n_\text{batch}}$ from $q$\;
    Compute the sampling PDF values $q(x_i)$\;
    Compute function values $f(x_i)$\;
    Start tracking gradients with respect to $\theta$\;
 \For{N steps}{
    Compute the PDF values from the parametric PDF $p(x_i,\theta)$\;
    Estimate the loss $\hat{L} = \displaystyle \frac{1}{N} \sum_{i=1}^{n_\text{batch}} \frac{f(x_i)^2}{p(x,\theta)q(x)}$\;
    Compute $\nabla_\theta \hat{L}$ using backpropagation\;
    Set $\theta \leftarrow \theta - \eta \nabla_\theta \hat{L} $\;
    Reset gradients\;
 }
 }
 \Return $p(x,\theta)$\;
 \caption{Backward sampling training in \zunis}
 \label{alg:training}
\end{algorithm}

\section{Fundamental limitations of the VEGAS algorithm}
\label{app:vegas_limits}
\input{appendix/vegas_problems.tex}

\FloatBarrier
\section{Supplementary Results}
\subsection{Qualitative examples}
\label{app:supp_figures}
\input{performance/supp_figs.tex}
\FloatBarrier
\newpage
\subsection{Systematic benchmark details}
\label{app:benchmark_details}
\input{appendix/benchmark_data.tex}
\FloatBarrier
\newpage
\subsection{Comparing \zunis with uniform sampling on matrix elements}
\label{app:flatME}
\input{appendix/flat_matrix_element.tex}
\FloatBarrier
\subsection{Effect of survey strategies}
\label{app:survey}
\input{appendix/survey_strategies.tex}
\FloatBarrier
\section{Exact minimzation of the neural importance sampling estimator variance}
\label{app:optimal_is}
\input{appendix/optimal_is.tex}
\section{High-level concepts of the \zunis API}
\subsection{Normalizing flows with \texttt{Flow} classes}
\label{sec:flow}
\input{framework/flows.tex}
\subsection{Training with the \texttt{Trainer} class}
\label{sec:trainer}
\input{framework/trainers.tex}
\subsection{Integrating with the \texttt{Integrator} class}
\label{sec:integrator}
\input{framework/integrators.tex}

\subsection{Implementing new coupling cells}
\label{app:new_cells}
\input{appendix/new_couplings.tex}
\section{Hardware setup details}
\label{app:hardware}
\input{appendix/hardware.tex}

\end{document}

%% file: header/header.tex
\input{header/packages.tex}
\input{header/macros.tex}
\input{header/meta_info.tex}
\input{header/settings.tex}


%% file: header/packages.tex
\usepackage{jheppub}
\usepackage{graphicx}
\usepackage{amsfonts}
\usepackage{amssymb}
\usepackage[colorlinks=true]{hyperref}
\usepackage{url} 
\usepackage[T1]{fontenc} 
\usepackage{xspace}
\usepackage{multirow}
\usepackage{placeins}
\usepackage{siunitx}
\usepackage{tikz}
\usepackage{mathtools}
\usepackage{bbm}
\usepackage[ruled,vlined,linesnumbered]{algorithm2e}
\usepackage{cleveref}
\usepackage{xcolor}
\usepackage{listings}
\usepackage[htt]{hyphenat}
\usepackage{caption}
\usepackage{subcaption}
\makeatletter
\gdef\@fpheader{}
\makeatother


%% file: header/macros.tex
\newcommand{\zunis}{\textsc{Z\"uNIS}\xspace}

\crefformat{footnote}{#2\footnotemark[#1]#3}


%% file: header/meta_info.tex
\title{Accelerating HEP simulations with Neural Importance Sampling}


%% file: header/settings.tex
\definecolor{dkgreen}{rgb}{0,0.6,0}
\definecolor{gray}{rgb}{0.5,0.5,0.5}
\definecolor{mauve}{rgb}{0.58,0,0.82}
\definecolor{mygray}{rgb}{0.9,0.9,0.9}
\definecolor{LightGray}{gray}{0.95}
\hypersetup{
  allcolors = [rgb]{0.1,0.1,0.5}
}


%% file: introduction/introduction.tex
High-Energy-Physics (HEP) simulations are at the heart of the Large Hadron Collider (LHC) program for studying the fundamental laws of nature.
Most HEP predictins are expressed as expectation values, evaluated numerically as Monte Carlo (MC) integrals. This permits both the integration of the very complex functions and the reproduction of the data selection process by experiments.

Most HEP simulations tools~\citep{madgraph, Reuter:2023vei,Sherpa:2019gpd} perform MC integrals using importance sampling, which allows to adaptively sample points to speed up convergence while keeping independent and identically distributed samples, crucial to reproduce experimental analyses which can only ingest uniformly-weighted data, typically produced by rejection sampling (see \cref{app:unweighting}).

The most popular tool to optimize importance sampling in HEP is by far the VEGAS algorithm~\citep{lepageVEGASADAPTIVEMULTIDIMENSIONAL1980}, which fights the curse of dimensionality by assuming no correlations between the variables. 
While this is rarely the case in general, a good understanding of the integrand function can help significantly.
Indeed optimized parametrizations using multichannelling~\citep{kleissNewMonteCarlo1986, Ohl:1998jn, kleissWeightOptimizationMultichannel1994} have become bread-and-butter tools for HEP event generation simulators, with good success for leading-order (LO) calculations.
However, as simulations get more complex, either by having more complex final states or by including higher orders in perturbation theory, performance degrades fast.

While newer approaches to adaptive importance sampling have been developed, for example Population Monte Carlo ~\citep{7974876,BUGALLO201536,cappe2004population, Iba:2000za} and its extensions~\citep{cappe2008adaptive, koblents2015population, ELVIRA201777,douc2007minimum, cornuet}, and have been successfully applied to other fields, VEGAS and its recent updates and variants \cite{Ohl:1998jn,Jadach:1999sf,Hahn:2004fe, vanHameren:2007pt, lepageAdaptiveMultidimensionalIntegration2021} remain one of the most commonly used algorithms in practically all modern simulation tools ~\citep{madgraph,Sherpa:2019gpd,Bellm:2015jjp}. For example, the WHIZARD event generator \cite{Reuter:2023vei} uses VAMP \cite{Ohl:1998jn}, which is an improvement of the orginal VEGAS implementation. 

We suspect that the reasons for this are twofold.
First of all the lack of exchange between the two scientific communities is probably at cause and should probably not to be neglected.
Second of all however, the impetus in the HEP community is to move toward more "black box" approaches where little is known of the structure of the integrand. This is one reason why approaches like PMC are less common, as it is sensitive to to the initial proposal density ~\citep{beaujean2013initializing, cappe2008adaptive}.
As the main practical goal of this paper is to reduce the reliance of HEP simulations on the careful tuning of integrators, we will focus on comparing our work with the \textit{de facto} HEP standard: the VEGAS algorithm. It is worth noting that alternative adaptive approaches, such as Nested Sampling \cite{10.1214/06-BA127, Handley_2015}, as employed in the SHERPA event generator \cite{Yallup:2022yxe}, also exist. However, for the sake of maintaining focus and as it is the most widespread tool, we will concentrate on comparing our work with VEGAS. Our comparison will be based on one of the most recent implementations \cite{lepageAdaptiveMultidimensionalIntegration2021}, which makes use of stratified sampling to boost the performance considerably.

There is much room for investing computational time into improving sampling~\citep{ATLASHLLHCComputing2020a}: modern HEP theoretical calculations are taking epic proportions and  can require hours for a single function evaluation~\citep{jonesHiggsBosonPair2018}. Furthermore, unweighting samples can be extremely inefficient, with upwards of 90\% sampled points discarded~\citep{foundationHLLHCComputingReview2020}. More powerful importance sampling algorithms would therefore be a welcome improvement~\citep{buckleyComputationalChallengesMC2020, wgChallengesMonteCarlo2021}.

Several approaches have been explored to tackle the challenge of efficient sampling in high-energy physics simulations. Initial efforts utilized classical neural networks for sampling, but often encountered prohibitive computational costs~\citep{bendavidEfficientMonteCarlo2017a, klimekNeuralNetworkBasedApproach2020, chenImprovedNeuralNetwork2021}.
 Another avenue of research has been the adoption of generative models, such as generative adversarial networks (GANs), which have shown promise in accelerating sampling speed significantly ~\citep{Butter_2019, Di_Sipio_2019, Butter_2020, Ahdida_2019, hashemi2019lhc, Carrazza_2019}. While such approaches do improve sampling speed by a large factor, they have major limitations. In particular, they have no theoretical guarantees of providing a correct answer on average \citep{bellagenteUnderstandingEventGenerationNetworks2021, Matchev:2020tbw}.

  On the other hand, tools like SHERPA offer an unbiased NN-powered approach for unweighting \cite{10.21468/SciPostPhys.12.5.164}. It is worth noting that reweighting techniques, such as the recently developed Deeply Conditionalized Transporter Networks (DCTR) \cite{Stoye:2018ovl,DiBello:2020ppq,diefenbacher2020dctrgan,andreassen2020neural}, have become standard in boosting the efficiency generated samples. Despite these advancements, drawbacks persist. For instance, the quality of information in the reweighted samples is restricted to the training sample \cite{bellagenteUnderstandingEventGenerationNetworks2021,Matchev:2020tbw,Butter:2020qhk}.

To avoid these disadvantages, our work exploits Neural Importance Sampling (NIS)~\citep{mullerNeuralImportanceSampling2019, zhengLearningImportanceSample2019}, which relies on normalizing flows and has strong theoretical guarantees.

A number of works have been published on using NIS for LHC simulations~\citep{gaoIflowHighdimensionalIntegration2020, bothmannExploringPhaseSpace2020a, gaoEventGenerationNormalizing2020,pina2020exhaustive,Heimel:2022wyj,heimel2023madnis}, as well as closely related variations~\citep{bellagenteUnderstandingEventGenerationNetworks2021, stienenPhaseSpaceSampling2021}, but most studies have focused on preliminary investigation of performance and less on the practical usability of the method.
Indeed, training requires function evaluations, which we are trying to minimize and data-efficiency training is therefore an important but often under-appreciated concern. Similarly, stability of training without the need of fine-tuning is essential for non-expert users. Furthermore, few authors have provided easily usable open source code, making the adoption of the technique in the HEP community difficult.

Our contribution to improve this situation can be summarized in three items:

\begin{itemize}
    \item The introduction of a new adaptive training algorithm for NIS. This permits the re-use of sampled points over multiple gradient descent steps, therefore making NIS much more data efficient, while at the same time guaranteeing stable and reliable training.
    \item A comparative study on the behaviour of different loss functions and their impact on performance for different processes.
    \item The introduction of \zunis, a \texttt{PyTorch}-based library providing robust and usable NIS tools, usable by non-experts. It implements previously-developped ideas as well as our new training procedure and is \href{https://zunis.readthedocs.io/en/stable/}{extensively documented}.
\end{itemize}


%% file: introduction/importance_sampling.tex
Importance sampling relies on the interpretation of integrals as expectation values. 
Indeed, let us consider an integral over a finite volume:
\begin{equation}
    I= \int_\Omega dx f(x),\quad \text{where } V(\Omega) = \int_\Omega dx \text{ is finite.}
\end{equation}
Let $p$ be a strictly positive probability distribution over $\Omega$, we can re-express our integral as the expectation of a random variable
\begin{align}
    I = \int_\Omega p(x) dx \frac{f(x)}{p(x)} = \underset{X_i\sim p}{\mathbb{E}} \frac{1}{N} \sum_{i=1}^N \frac{f(X_i)}{p(X_i)},
\end{align}
whose mean is indeed $I$ and whose standard deviation is $\dfrac{\sigma(f,p)}{\sqrt{N}}$, where $\sigma(f,p)$ is the standard deviation of $f(X)/p(X)$ for $X\sim p$:
\begin{equation}
    \sigma^2(f,p) = \underset{x\sim p}{\mathbb{E}}\hspace{-.1em} \left(\left(\frac{f(x)}{p(x)}\right)^2\right) - I^2.
\end{equation}
The problem statement of importance sampling is to find the probability distribution function $p$ that minimizes the variance of our estimator for a given $N$. 
In the case of multichanneling \cite{kleissWeightOptimizationMultichannel1994,Weinzierl:2000wd}, finding the probability distribution function $p$ is replaced by finding weights $\alpha_i$ and mappings $g_i$ such that
\begin{equation}
    g(x)=\sum_{i}^m \alpha_i g_i(x), \quad \sum_i^{m}\alpha_i=1 
\end{equation}
and
\begin{equation}
    I=\sum_i^m \int dx \alpha_i g_i(x) \frac{f(x)}{g(x)}
\end{equation}
The $g_i$ are often chosen according to prior physics knowledge. As we are interested in a blackbox approach, we do not use different channels but instead try to learn an optimal probability distribution function. 
In Neural Importance Sampling, we rely on Normalizing Flows to approximate the optimal distribution, which we can optimize using stochastic gradient descent.
However, it has been shown that both approaches can be successfully combined \cite{Heimel:2022wyj, heimel2023madnis}. Nevertheless, in this work we focus on achieving improvements with a single channel, in order to minimize the required prior knowldge of the integrand.


%% file: introduction/normalizing_flows.tex
Normalizing flows~\citep{tabak2010density,tabak2013family,rippel2013high,
 rezendeVariationalInferenceNormalizing2015} provide a way to generate complex probability distribution functions from simpler ones using parametric changes
of variables that can be learned to approximate a target distribution. 
As such, normalizing flows are diffeomorphisms: invertible, (nearly-everywhere) differentiable mappings with a differentiable inverse. 

Indeed, if $u \sim p(u)$, then $T(u) = x \sim q(x)$ where
\begin{equation}
    q\left( x = T(u)\right) = p(u) \left| J_T \right(u)|^{-1},
\end{equation}
where $J_T$ is the Jacobian determinant of $T$:
\begin{equation}
    J_T(u) = \det \frac{\partial T_i}{\partial u_j} (u).
\end{equation}

In the world of machine learning, $T$ is called a normalizing flow and is typically part of a parametric family of diffeomorphisms ($T(\cdot, \theta)$) such that gradients $\nabla_\theta J_T$ are tractable.

Coupling cell mappings perfectly satisfy this requirement~\citep{dinh2014nice, dinh2016density, mller2018neural}: they are neural-network-parametrized bijections whose Jacobian factor can be obtained analytically without backpropagation or expensive determinant calculation.
As such, they provide a good candidate for importance sampling as long as they can be trained to learn an unnormalized target function, which is exactly what neural importance sampling proposes.

The coupling cells used in the following implement the transformations proposed in ~\citep{dinh2014nice, dinh2016density, mller2018neural}, which are also used in i-flow and MadNIS. Although all the transformations are implemented, we focus our study on the use piecewise-quadratic layers, as these reach a high level of expressiveness without requiring a too large number of parameters. This is important, as a too high number of transformation parameters becomes expensive to learn. 

Other than the architecture of NNs used by MadNIS in \cite{Heimel:2022wyj} and also the architecture chosen by i-flow, we use by default considerably larger NN (rectangular DNN with 8 hidden layers, each with 256 nodes), as the we learn the probability distribution function directly instead of splitting it into multiple channels. This guarantees also enough complexity in order to allow stable training on a wide range of integrands. Although we also use LeakyReLU activation functions, we additionally add an input activation layer which scales the input and removes any offset to increase stability of trainging. Additionally, we implement the optimal masking function proposed by i-flow \cite{gaoIflowHighdimensionalIntegration2020}.


%% file: introduction/neural_importance_sampling.tex
Neural importance sampling was introduced in the context of computer graphics~\citep{mller2018neural} and proposes to use normalizing flows as a family of probability distributions over which to solve the minimization problem of importance sampling.

\begin{equation}
    \label{eq:VarianceLossIntegralRecall}
    \mathcal{L}(\theta) = \int_\Omega dx \frac{f^2(x)}{p\left(x,\theta\right)}.
\end{equation}
Of course, to actually do so, one needs to find a way to explicitly evaluate $\mathcal{L}(\theta)$ and the original neural importance sampling approach proposes to approximate it using importance sampling. 
One needs to be careful that the gradient of the estimator of the loss need not be the estimator of the gradient of the loss. 
The gradient of the loss can be expressed as
\begin{equation}
    \nabla_\theta \mathcal{L}(\theta) = -\int_\Omega dx \frac{f^2(x)}{p\left(x,\theta\right)} \nabla_\theta \log p\left(x,\theta\right),
\end{equation}
for which an estimator is proposed as
\begin{equation}
\label{eq:NISLossEstimator}
    \widehat{\nabla}_\theta \mathcal{L}(\theta) = - \sum_{i=0}^N \left(\frac{f(X_i)}{p\left(X_i,\theta\right)}\right)^2 \nabla_\theta \log p\left(X_i,\theta\right),\quad X_i \sim p.
\end{equation}
The authors also observed that other loss functions are possible which share the same global minimum as the variance based loss: for example, the Kullback-Leibler divergence $D_{KL}$ between two functions is also minimized when they are equal.
Such alternative loss functions are not guaranteed to work for importance sampling, but they prove quite successful in practice. Indeed, for certain processes they are shown to be crucial for optimal performance, as can be seen in section \ref{sec:madgraph}.
After training to minimize the loss estimator of \cref{eq:NISLossEstimator}, the normalizing flows provides a tractable probability distribution function from which to sample points and estimate the integral.


%% file: framework/training.tex
In this section, we describe how we train probability distributions within \zunis using gradient-based optimizers.
While the solution proposed in the original formulation of NIS defined \cref{eq:NISLossEstimator} works and has been successfully used by i-flow, its main drawback is that it samples points from the same distribution that it tries to optimize. As a result, new points $X_i$ must be sampled from $p$ after every gradient step, which is very inefficient for slow integrands.

Our solution to this problem is to remember that the loss function is an integral, which can be evaluated by importance sampling using any PDF, not only $p$. We will therefore define an auxiliary probability distribution function $q(x)$, independent from $\theta$, from which we sample to estimate our loss function:
\begin{equation}
    \int dx \frac{f(x)^2}{p(x,\theta)} = \underset{x\sim q}{\mathbb{E}} \frac{f(x)^2}{p(x,\theta) q(x)}.
\end{equation}

This is the basis for the general method we use for training probability distributions within \zunis, described in \cref{alg:training}. Because the sampling distribution is separated from the model to train, the same point sample can be reused for multiple training steps, which is not possible when using \cref{eq:NISLossEstimator}. This approach is similiar to the "buffered training" used in MadNIS \cite{Heimel:2022wyj}. This is particularly important for high-precision particle physics predictions that involve high-order perturbative calculations or complex detector simulations because function evaluations can be extremely costly. We show in \cref{sec:performance}, in particular in \cref{fig:camel_benchmark_epochs} that reusing data indeed has a very significant impact on data efficiency.

\begin{algorithm}
    \SetAlgoLined
    \KwData{Parametric PDF $p(x,\theta)$}
    \KwResult{Trained PDF $p(x,\theta)$}
     \For{M steps}{
        Sample a batch $x_1,\dots,x_{n_\text{batch}}$ from $q$\;
        Compute the sampling PDF values $q(x_i)$\;
        Compute function values $f(x_i)$\;
        Start tracking gradients with respect to $\theta$\;
     \For{N steps}{
        Compute the PDF values from the parametric PDF $p(x_i,\theta)$\;
        Estimate the loss $\hat{L} = \displaystyle \frac{1}{N} \sum_{i=1}^{n_\text{batch}} \frac{f(x_i)^2}{p(x,\theta)q(x)}$\;
        Compute $\nabla_\theta \hat{L}$ using backpropagation\; \label{alg:training:gradient}
        Set $\theta \leftarrow \theta - \eta \nabla_\theta \hat{L} $\;
        Reset gradients\;
     }
     }
     \Return $p(x,\theta)$\;
     \caption{Backward sampling training in \zunis}
     \label{alg:training}
    \end{algorithm}

After training, $q$ is discarded and the integral is estimated from the optimized $p$. 

The only constraint on $q$ is that it yields a good enough estimate so that gradient steps improve the model. Much like in other applications of neural networks, we have found that stochastic gradient descent can yield good results despite noisy loss estimates. We propose three schemes for $q$:
\begin{itemize}
    \item A uniform distribution (\lstinline{survey_strategy="flat"})
    \item A frozen copy of the model, which can be updated once in a while\footnote{This is inspired by deep-$Q$ learning, where two copies of the value model are used: a frozen one used to sample actions, and a trainable one used to estimate values in the loss function. Here the frozen copy is used to sample points, and the trainable model is used to compute PDF values used in the loss function} (\lstinline{survey_strategy="forward"})
    \item An adaptive scheme starting with a uniform distribution and switching to sampling from a frozen model when it is expected to yield a more accurate loss estimate (\lstinline{survey_strategy="adaptive_variance"}).
\end{itemize}

An important point to notice is that the original NIS formulation in~\cref{eq:NISLossEstimator} can be restated as a limiting case of our approach. Indeed, if we take $q$ to be a frozen version of the model $p(x,\theta_0)$, which we update everytime we sample points (setting $N=1$ in \cref{alg:training}), the gradient update in \cref{alg:training:gradient} is
\begin{align}
    \left.\nabla_\theta \left[\underset{x\sim p(x,\theta_0)}{\mathbb{E}} \frac{f(x)^2}{p(x,\theta) p(x,\theta_0)}\right]\right\rvert_{\theta_0 = \theta} = - \underset{x\sim p(x,\theta)}{\mathbb{E}} \frac{f(x)^2}{p(x,\theta)^2} \nabla_\theta \log p(x,\theta).
    \label{eq:new_to_old_nis}
\end{align}
The core difference to the "buffered training" approach presented by MadNIS is that we offer an adaptive scheme, which automatically switches between which distribution is chosen to sample from. Using a uniform distribution in the beginning ensures that independent of initialisation every relevant region of the integration space is covered. However, with progressing training the frozen model becomes more desirable. In the adaptive training approaches, we switch once both losses become comparable. This is a crucial improvement, as it allows more stable training even without prior knowledge of the integrand. Additionally, a good coverage at early times of the training can speed up the training process.

%% file: framework/framework.tex
On the practical side, \zunis is a \texttt{PyTorch}-based library which implements many ideas formulated in previous work but organizes them in the form of a modular software library with an easy-to-use interface and well-defined building blocks. We believe this structure will help non-specialist use it without understanding all the nuts and bolts, while experts can easily add new features to responds to their needs.
The \zunis library relies on three layers of abstractions which steer the different aspects of using normalizing flows to learn probability distributions from un-normalized functions and compute their integrals: 
\begin{itemize}
    \item \texttt{Flows}, which implement a bijective mapping which transforms points and computes the corresponding Jacobian are described in \cref{sec:flow}
    \item \texttt{Trainers}, which provide the infrastructure to perform training steps and sample from flow models are described in \cref{sec:trainer}
    \item \texttt{Integrators}, which use trainers to steer the training of a model and compute integrals are described in \cref{sec:integrator}
\end{itemize}

%% file: performance/perf.tex
In this section, we evaluate \zunis on a variety of test functions to assess its performance and compare it to the commonly used VEGAS algorithm~\citep{PETERLEPAGE1978192,Ohl:1998jn}. Although there have been recent advances in computing both matrix elements and the integration using VEGAS on GPU \cite{bothmann2023portable}, we focus on the predominant case of these steps being performed on CPU.
We first produce a few low dimensional examples for illustrative purposes, then move on to integrating parametric functions in various dimensions and finally evaluate performance on particle scattering matrix elements.


%% file: performance/qualitative.tex
Let us start by illustrating the effectiveness of \zunis in a low dimensional setting where we can readily visualize results. We define three functions on the 2D unit hypercube which each illustrate some failure mode of VEGAS (see~\cref{app:vegas_limits}).

We ran the \zunis \texttt{Integrator} with default arguments over ten repetitions for each function and report the performance of the trained integral estimator compared to a flat estimator and to VEGAS in \cref{tab:2dresults}. Overall, \zunis \texttt{Integrators} learn to sample from their target function extremely well: we outperform VEGAS by a factor 100 for the camel and the slashed circle functions and a factor 30 for the sinusoidal function and VEGAS itself provides no advantage over uniform sampling for the latter two.

\begin{table}[!ht]
    \centering
    \begin{tabular}{c|ccc}
        \hline
        Variance Reduction & Camel & Slashed Circle & Sinusoidal \\\hline
        vs. uniform & $1.8 \pm 0.4 \times 10^3$ & $8.9 \pm 0.9 \times 10^1$ & $2.0 \pm 0.5 \times 10^2$ \\
        vs. VEGAS & $7.0 \pm 1.4 \times 10^2$ & $8.8 \pm 0.9 \times 10^1$  & $1.6 \pm 0.5 \times 10^2$ \\
    \end{tabular}
    \caption{Variance reduction (high is good) for the camel, slashed circle and sinusoidal functions compared to uniform sampling and to VEGAS over 10 repetitions.}
    \label{tab:2dresults}
\end{table}

\begin{figure}[!ht]
    \centering
    \begin{subfigure}[b]{0.3\textwidth}
        \centering
        \includegraphics[height=8cm,trim={0 11.8cm 0 0},clip]{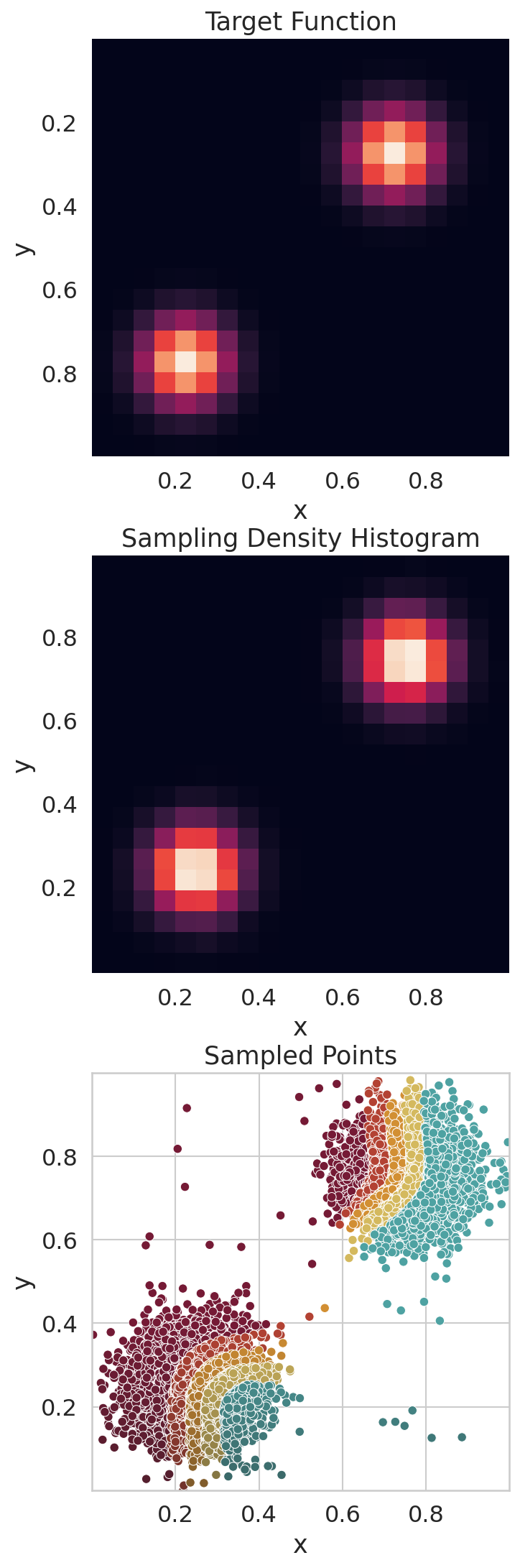}
        \caption{Camel function}
        \label{fig:2dcomparisoncamel}
    \end{subfigure}
    \begin{subfigure}[b]{0.3\textwidth}
        \centering
        \includegraphics[height=8cm,trim={0 11.8cm 0 0},clip]{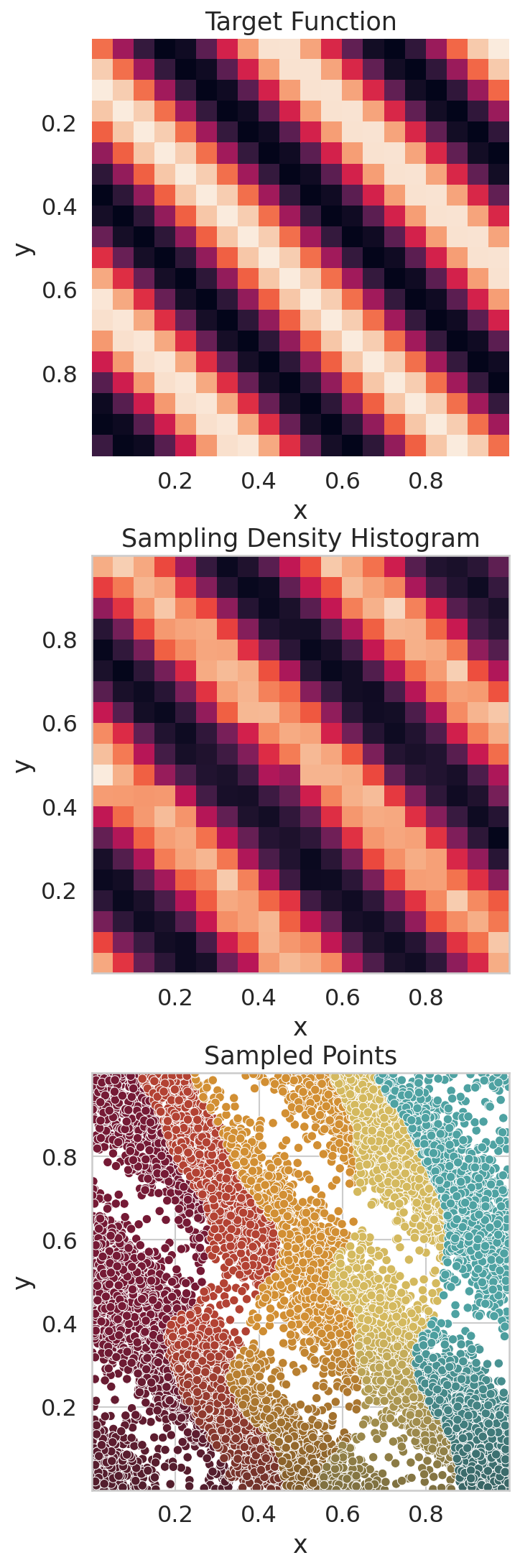}
        \caption{Sinusoidal function}
        \label{fig:2dcomparisonwave}
    \end{subfigure}
    \begin{subfigure}[b]{0.3\textwidth}
        \centering
        \includegraphics[height=8cm,trim={0 11.8cm 0 0},clip]{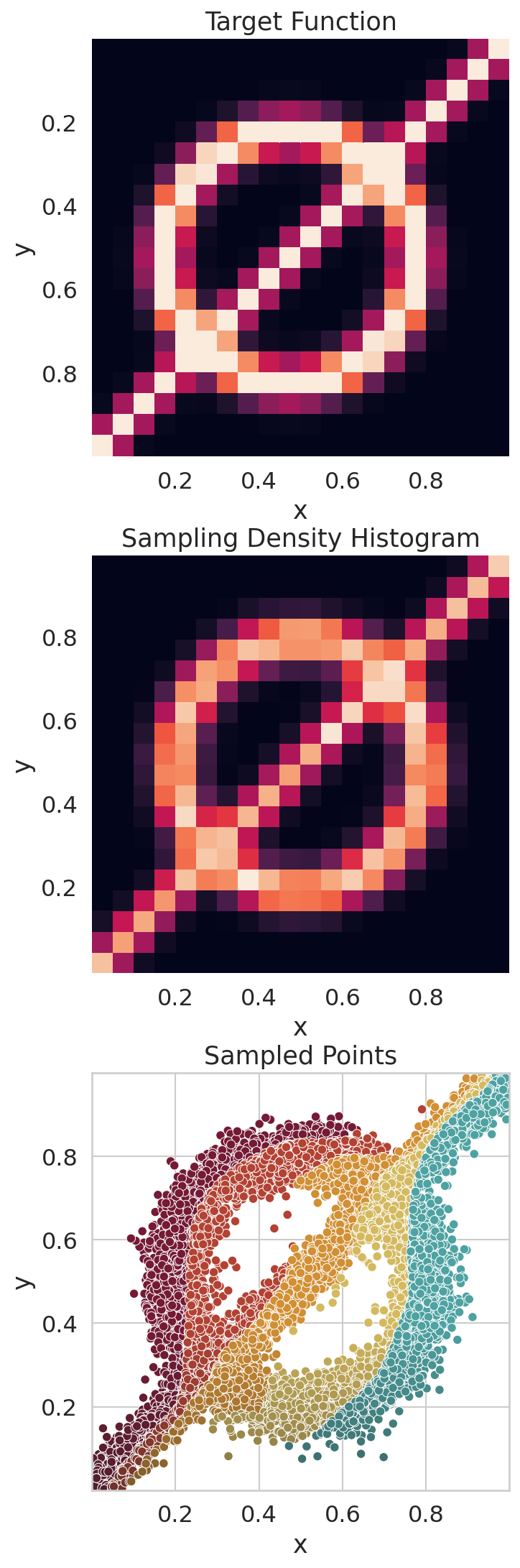}
        \caption{Slashed circle function}
        \label{fig:2dcomparisoncircle}
    \end{subfigure}
    \caption{Comparison between target functions and point sampling densities for \ref{fig:2dcomparisoncamel} the camel function, \ref{fig:2dcomparisonwave} the sinusoidal function, \ref{fig:2dcomparisoncircle} the slashed circle function. Supplementary \cref{fig:2dmapping} shows how points are mapped from latent to target space.}
    \label{fig:2dcomparison}
\end{figure}

We further illustrate the performance of our trained models by comparing the target functions and density histogram for points sampled from the normalizing flows in \cref{fig:2dcomparison}, which shows great qualitative agreement. 

%% file: performance/benchmarks.tex
Let us now take a more systematic approach to benchmarking \zunis. We compare \zunis \hyperref[sec:integrator]{Integrators} against uniform integration and VEGAS using the following metrics:
integrand variance (a measure of convergence speed, see \cref{sec:importance_sampling}), unweighting efficiency (a measure of the efficiency of exact sampling with rejection, see \cref{app:unweighting}) and wall-clock training and sampling\footnote{We provide details on hardware in \cref{app:hardware}\label{foot:hardware}}.

\paragraph{\zunis improves convergence rate compared to VEGAS}
For this experiment, we focus on the camel function defined in \cref{eq:came_def} and scan a 35 configurations spanning from $2$ to $32$ dimensions over function variances between $10^{-2}$ and $10^{2}$ as shown in \cref{tab:camel_setups}.

\begin{figure}[!ht]
    \centering
    \begin{subfigure}[b]{0.49\textwidth}
        \includegraphics[width=\textwidth]{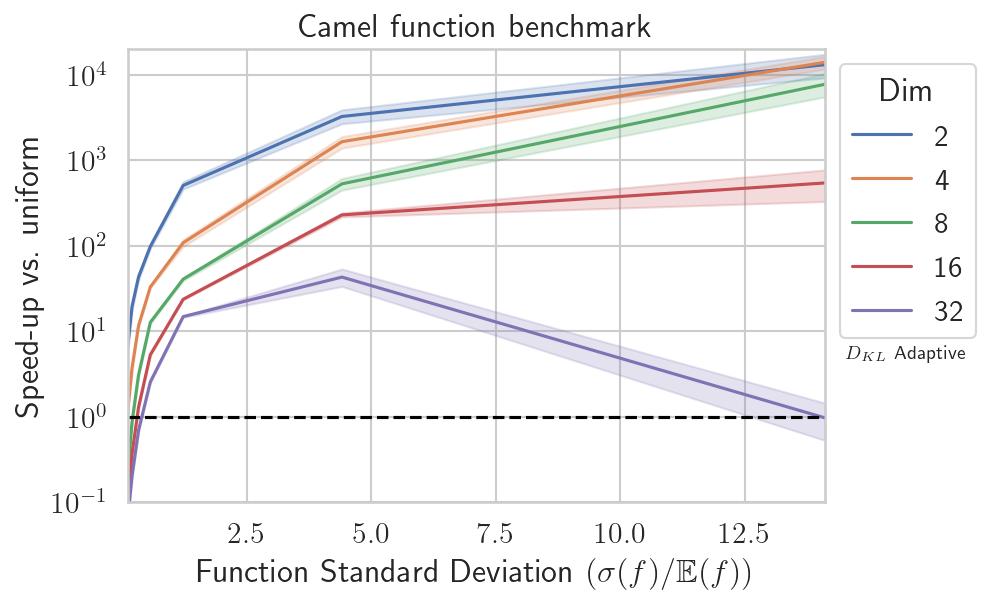}
        \caption{}
        \label{fig:camel_benchmark_default_uniform_sd}
    \end{subfigure}
    \begin{subfigure}[b]{0.49\textwidth}
        \includegraphics[width=\textwidth]{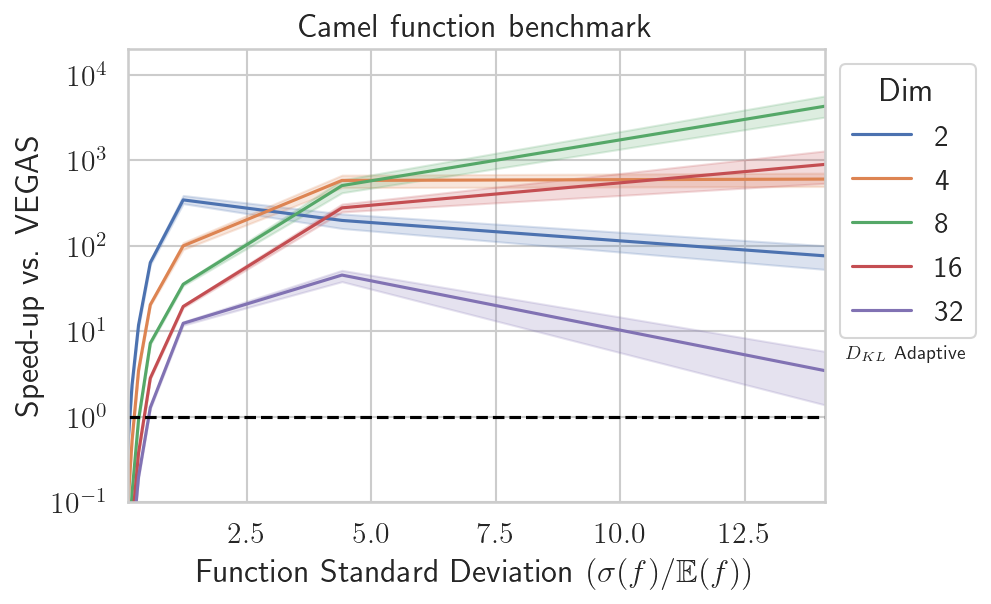}
        \caption{}
        \label{fig:camel_benchmark_default_vegas_sd}
    \end{subfigure}
    \begin{subfigure}[b]{0.49\textwidth}
        \includegraphics[width=\textwidth]{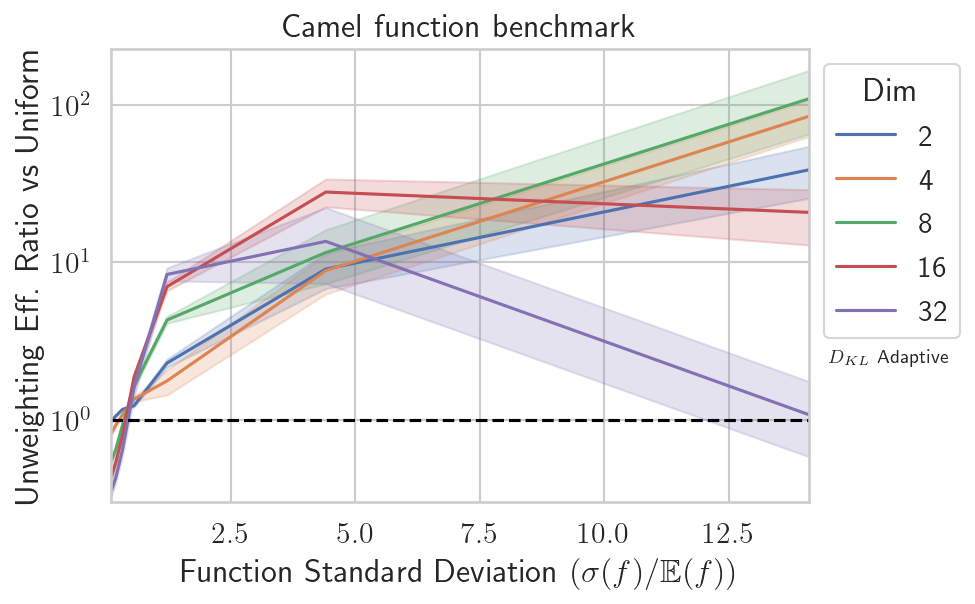}
        \caption{}
        \label{fig:camel_benchmark_default_uniform_unw}
    \end{subfigure}
    \begin{subfigure}[b]{0.49\textwidth}
        \includegraphics[width=\textwidth]{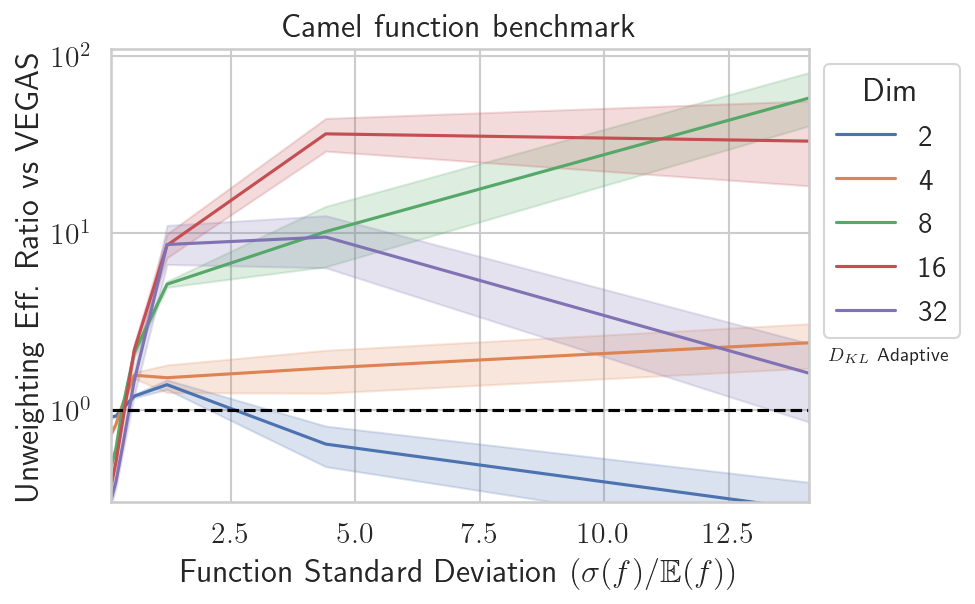}
        \caption{}
        \label{fig:camel_benchmark_default_vegas_unw}
    \end{subfigure}
    \caption{Benchmarking \zunis against uniform sampling and VEGAS  with default settings. 
     In (\ref{fig:camel_benchmark_default_uniform_sd}-\ref{fig:camel_benchmark_default_vegas_sd}), we show the sampling speed-up (ratio of integrand variance) as a function of the relative standard deviation of the integrand, while we show the unweighting speed-up (ratio of unweighting efficiencies) in (\ref{fig:camel_benchmark_default_uniform_unw}-\ref{fig:camel_benchmark_default_vegas_unw}).}
    \label{fig:camel_benchmark_default}
\end{figure}

Except in the low variance limit, \zunis can reduce the required number of points sampled to attain a given precision on integral estimates without any parameter tuning, attaining speed-ups of up to $\times 1000$ both compared to uniform sampling and VEGAS-based importance sampling, as shown in \cref{fig:camel_benchmark_default_uniform_sd}-\ref{fig:camel_benchmark_default_vegas_sd} and \cref{tab:camel_default_results}. 
Compared with the results reported in Table II of \cite{gaoIflowHighdimensionalIntegration2020}, we see also a significant improvement with respect to i-flow. i-flow reports only moderate reduction of needed function calls up to one third, whereas we find in the same range of dimensions and integrand variances (multiple) order of magnitude reductions in variance, which shows the strength of our approach.
Unweighting efficiencies are also boosted significantly, although more mildly than variances, as shown in \cref{fig:camel_benchmark_default_uniform_unw}-\ref{fig:camel_benchmark_default_vegas_unw}, which we could attribute to PDF underestimation in regions with low point density; the nature of the veto algorithm makes it very sensitive to a few bad behaving points in the whole dataset.

\begin{figure}[!ht]
    \centering
    \begin{subfigure}[b]{0.49\textwidth}
    \includegraphics[height=.61\textwidth]{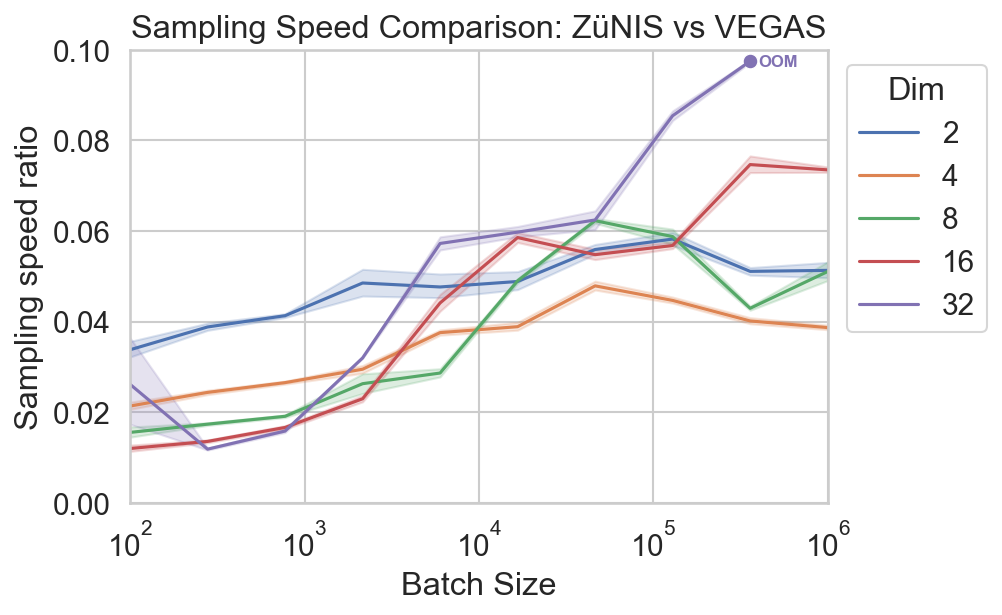}
    \caption{}
    \label{fig:camel_benchmark_default_vegas_speed}
    \end{subfigure}
    \begin{subfigure}[b]{0.49\textwidth}
    \includegraphics[height=.61\textwidth]{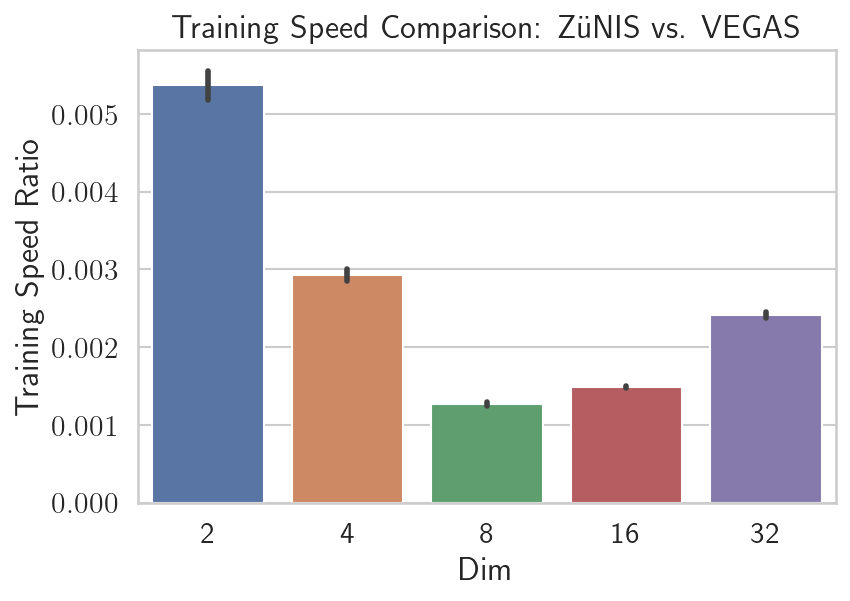}
    \caption{}
    \label{fig:camel_benchmark_default_vegas_train_speed}
  \end{subfigure}
  
    \caption{Comparison of the training and sampling speed of \zunis and VEGAS. As can be expected, \zunis is much slower than VEGAS, both for training and sampling, although larger batch sizes can better leverage the power of hardware accelerators.}
    \label{fig:camel_benchmark_speeds}
\end{figure}

\paragraph{\zunis is slower than VEGAS}
\zunis does not, however, outclass VEGAS on all metrics by far: as shown in \cref{fig:camel_benchmark_speeds}, training is a few hundred times slower than VEGAS and sampling is 10-50 times slower, all while \zunis runs on GPUs. This is to be expected given the much increased computational complexity of normalizing flows compared to the VEGAS algorithm. As such, \zunis is not a general replacement for VEGAS, but provides a clear advantage for integrating time-intensive functions, where sampling is a negligible overhead, such as precision high-energy-physics simulations.

\begin{figure}[!ht]
    \centering
    \begin{subfigure}[b]{0.49\textwidth}
    \includegraphics[height=.61\textwidth]{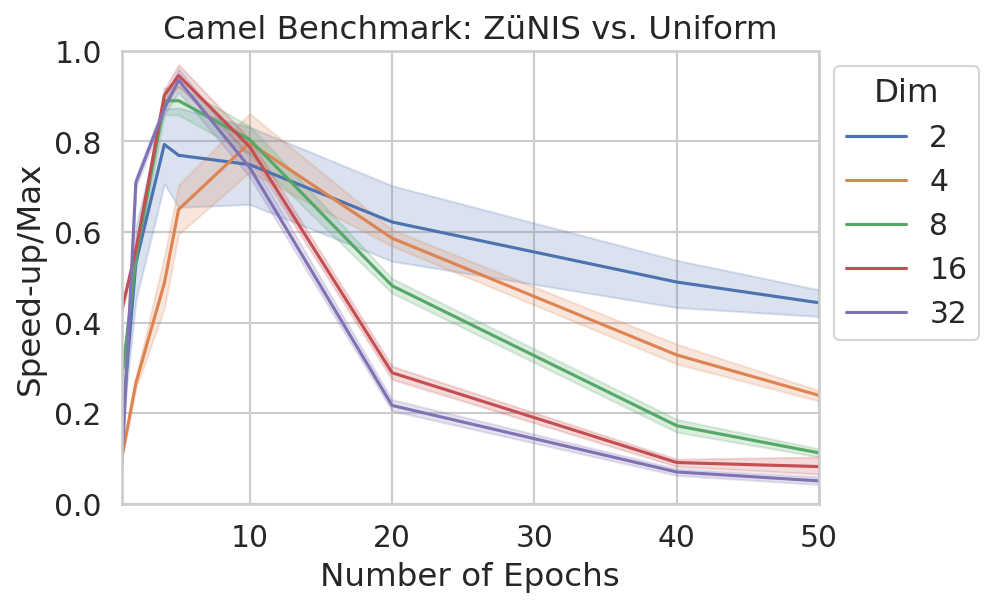}
    \caption{}
    \label{fig:camel_benchmark_default_uniform_epochs}
    \end{subfigure}
    \begin{subfigure}[b]{0.49\textwidth}
    \includegraphics[height=.61\textwidth]{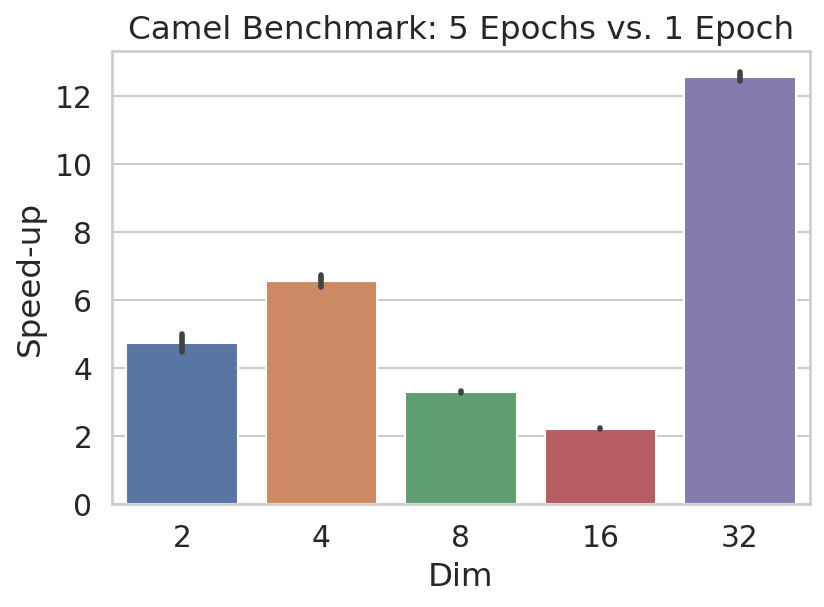}
    \caption{}
    \label{fig:camel_benchmark_default_5v1_epochs}
    \end{subfigure}
    
    \caption{\Cref{fig:camel_benchmark_default_uniform_epochs}: Effect of repeatedly training on the same sample of points over multiple epochs. For all settings, there is a large improvement when going from one to moderate epoch counts, with a peak around 5-10. Larger number of epochs lead to overfitting, which impacts performance negatively. \Cref{fig:camel_benchmark_default_5v1_epochs}: Comparison between optimal data reuse (5 epochs) and the original NIS algorithm (1 epoch).}
    \label{fig:camel_benchmark_epochs}
\end{figure}

\paragraph{The new loss function introduced in \zunis improves data efficiency}
We have shown that \zunis is a very performant importance sampling and event generation tool and provides significant improvements over existing tools, while requiring little fine tuning from users. Another key result is that the new approach to training we introduced in \cref{sec:training} has a large positive impact on performance. Indeed, as can be seen in \cref{fig:camel_benchmark_epochs}, re-using samples for training over multiple epochs provides a 2- to 10-fold increase in convergence speed, making training much more data-efficient. 

For this experiment, we use forward sampling, where the frozen model is used to sample a batch of points which are then used for training over multiple epochs before resampling from an update of the frozen model.
As a result, we reproduce the original formulation of NIS in~\cref{eq:NISLossEstimator} when we use a single epoch as shown in \cref{eq:new_to_old_nis}.

%% file: performance/madgraph_iclr.tex
Cross-sections are integrals of quantum transition matrix elements for a a scattering process such as a LHC collision and express the probability that specific particles are produced. Matrix elements themselves are un-normalized probability distributions for the configuration of the outgoing particles: it is therefore both valuable to integrate them to know the overall frequency of a given scattering process, and to sample from them to understand how particles will be spatially distributed as they fly off the collision point.

\begin{figure}[!ht]
    \centering
    \begin{subfigure}[b]{.3\textwidth}
        \centering
        \includegraphics[width=\textwidth]{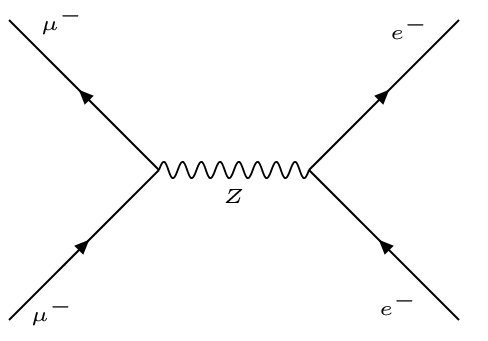}
        \caption{}
    \end{subfigure}
    \begin{subfigure}[b]{.3\textwidth}
        \centering
        \includegraphics[width=\textwidth]{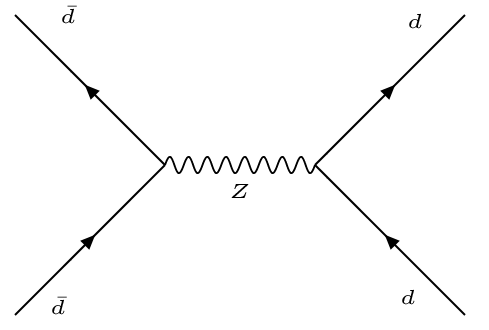}
        \caption{}
    \end{subfigure}
    \begin{subfigure}[b]{0.3\textwidth}
        \centering
        \includegraphics[width=\textwidth]{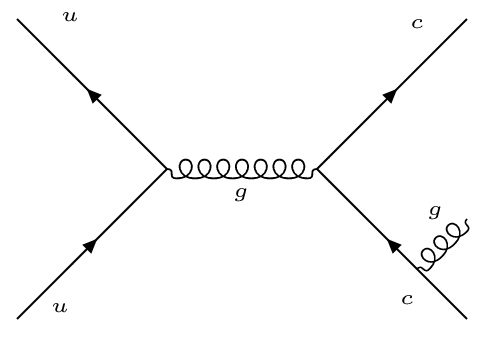}
        \caption{}
    \end{subfigure}
    \caption{Sample Feynman Diagrams for $ e^- \mu \rightarrow e^- \mu$ via $Z$, $ d \bar d \rightarrow d \bar d$ via $Z$ and $ u c \rightarrow u c g$ .}
    \label{fig:feyn}
\end{figure}

We study the performance of \zunis in comparison to \textsc{VEGAS} by studying three simple processes at leading order in perturbation theory, $ e^- \mu \rightarrow e^- \mu$ via $Z$, $ d \bar d \rightarrow d \bar d$ via $Z$ and $ u c \rightarrow u c g$ (with 3-jet cuts based on $\Delta R$), see \cref{tab:processes} and \cref{fig:feyn}. We use the first process as a very easy reference while the two other, quark-initiated processes are used to illustrate specific points. Indeed, both feature narrow regions of their integration space with large enhancements, due respectively to $Z$-boson resonances and infra-red divergences.

\begin{table}[!ht]
    \centering
    \begin{tabular}{c|ccc}
        \hline
         &  $ e^- \mu \rightarrow e^- \mu$ via $Z$ & $ d \bar d \rightarrow d \bar d$ via $Z$ & $ u c \rightarrow u c g$ \\\hline
        dimensions & 2 & 4 & 7\\
        normalized standard deviation & $1.45 \times 10^{-2}$ & $6.57 \times 10^{-2}$  & $0.96$ \\
    \end{tabular}
    \caption{Comparison of the three test processes.}
    \label{tab:processes}
\end{table}

We evaluate the matrix elements for these three processes by using the \textsc{Fortran} standalone interface of \textsc{MadGraph5\_aMC@NLO} \citep{madgraph}. The two hadronic processes are convolved with parton-distribution functions from \texttt{LHAPDF6} \citep{lhapdf}. We parametrize phase space (the particle configuration space) using the RAMBO on diet algorithm~\citep{pltzer2013rambo} implemented for \texttt{PyTorch} in \textsc{TorchPS} \citep{niklas_gotz_2021_4639109}.
Using RAMBO on diet has the advantage that no knowledge about the resonance structure of the integrand is required. However, this comes at a cost of reduced performance, as resonances are not reflected appropriately in the phase space sampling. It has been shown this can have a substantial effect on performance \cite{10.21468/SciPostPhys.15.4.169}.

\begin{figure}[!ht]
    \centering
    \begin{subfigure}[b]{0.49\textwidth}
        \centering
        \includegraphics[width=\textwidth]{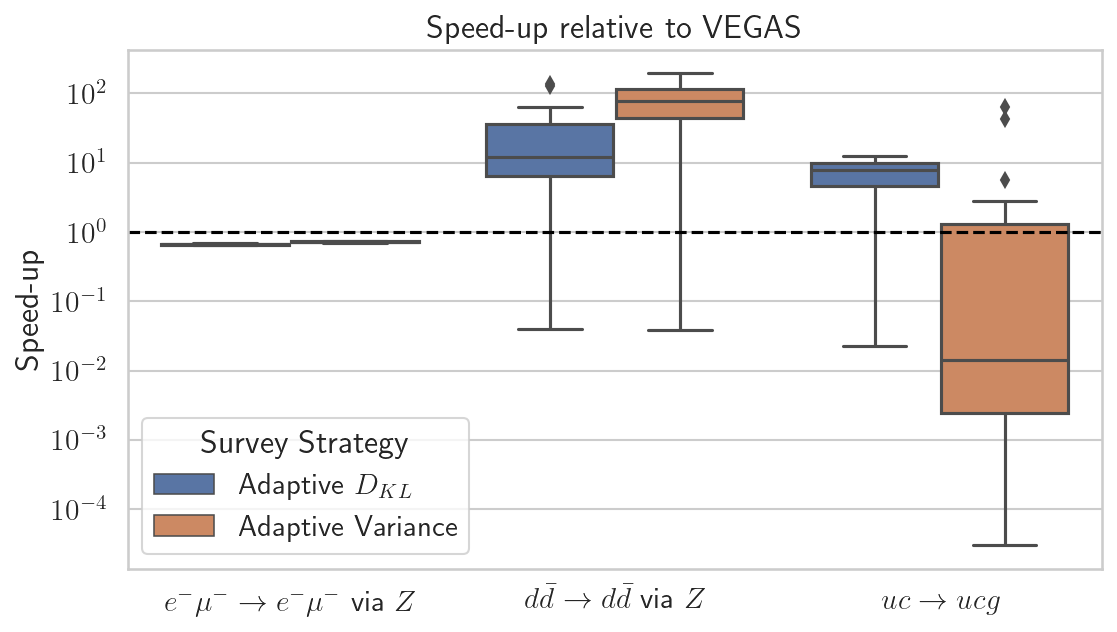}
        \caption{}
        \label{fig:madgraph_comparison_speed}
    \end{subfigure}
    \begin{subfigure}[b]{0.49\textwidth}
        \centering
        \includegraphics[width=\textwidth]{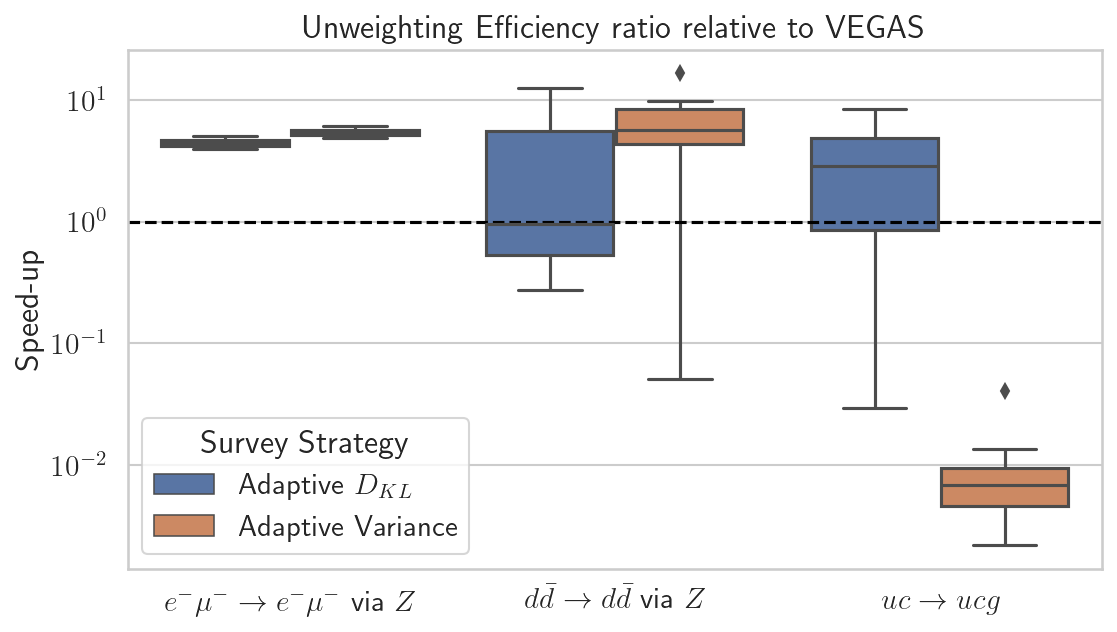}
        \caption{}
        \label{fig:madgraph_comparison_uwr}
    \end{subfigure}
    \caption{Average performance of \zunis over 20 runs relative to VEGAS, measured by the relative speed-up in \cref{fig:madgraph_comparison_speed} and by the relative unweighting efficiency in \cref{fig:madgraph_comparison_uwr}.}
    \label{fig:madgraph_comparison}
\end{figure}

We report benchmark results in \cref{fig:madgraph_comparison}, in which we trained over 500,000 points for each process using near-default configuration, scanning only over variance and Kullback-Leibler losses.

As previously observed, little convergence acceleration is achieved for low variance integrands like $ e^- \mu \rightarrow e^- \mu$, but unweighting still benefits from NIS. The two hadronic processes illustrate typical features for cross sections: training performance is variable and different processes are optimized by different loss function choices\footnote{Generally, smoother functions are better optimized with the Kullback-Leibler loss while functions with peaks benefit from using the variance loss. As we show in \cref{app:survey}, choosing an adaptive strategy is generally advisable whatever the loss}.

The performance of $d\bar d\to d\bar d$ shows nice improvement with \zunis while that of $uc\to ucg$ is more limited. This can be understood by comparing to importance sampling (see \cref{app:flatME}): it is in fact VEGAS, which performs significantly better on $uc\to ucg$ compared to $d\bar d\to d\bar d$ because the parametrization of RAMBO is based on splitting invariant masses, making them aligned with the enhancements in the $ucg$ phase space and allowing great VEGAS performance. This drives a key conclusion for the potential role of \zunis in the HEP simulation landscape: not to replace VEGAS, but to fill in the gaps where it fails due to inadequate parametrizations, as we illustrate here by using non-multichanneled $d \bar d \to d \bar d$ as a proxy for more complex processes.


%% file: conclusion/conclusion.tex
We have showed that \zunis can outperform VEGAS both in terms of integral convergence rate and unweighting efficiency on specific cases, at the cost of a significant increase in training and sampling time, which is an acceptable tradeoff for high-precision HEP computations with high costs. In this context, the introduction of efficient training is a key element to making the most of the power of neural importance sampling where function evaluation costs are a major concern. While further testing is required to ascertain how far NIS can fill the gaps left by VEGAS for integrating complex functions, there is already good evidence that \zunis can provide needed improvements in specific cases. We hope that the publication of a usable toolbox for NIS such as \zunis will stir a wider audience within the HEP community to apply the method so that the exact boundaries its applicability can be more clearly ascertained.


%% file: conclusion/acknowledgements.tex
We would like to thank Simone Lionetti, Armin Schweizer and Valentin Hirschi for many useful discussions. We are very grateful to Prof. Babis Anastasiou for his continued support providing computational resources.
N.D. received funding from ERC grand 69471 at ETHZ and is supported at IBM by SNF grant 200021\_192128 / 1.  N.G. acknowledges support by the Stiftung Polytechnische Gesellschaft Frankfurt am Main as well as the Studienstiftung des Deutschen Volkes, as well as by the Deutsche Forschungsgemeinschaft (DFG, German Research Foundation) – Project number 315477589 – TRR 211.

%% file: conclusion/reproducibility.tex
The library is available on \href{https://github.com/ndeutschmann/zunis/}{Github} or at PyPI \lstinline{pip install zunis}.

The recommended procedure to reproduce the experiments is to clone the repository and install the \texttt{Python} requirements using \lstinline{pip install -r requirements.txt}.

The data to reproduce the experiments can be generated using scripts provided in the repository at \lstinline{experiments/benchmarks}, in which \texttt{Jupyter} notebooks are also available to reproduce the figures of the paper. The following scripts are available:
\begin{itemize}
  \item \lstinline{benchmarks_03/camel/run_benchmark_defaults.sh} to generate camel integration data
  \item \lstinline{benchmarks_04/camel/run_benchmark_defaults.sh} to generate camel sampling speed data
  \item \lstinline{benchmark_madgraph/ex_benchmark_emu.sh} to generate  $ e^- \mu \rightarrow e^- \mu$ via $Z$ integration data
  \item \lstinline{benchmark_madgraph/ex_benchmark_dd.sh} to generate  $ d \bar d \rightarrow d \bar d$ via $Z$integration data
  \item \lstinline{benchmark_madgraph/ex_benchmark_ucg.sh} to generate  $uc\to ucg$ integration data
\end{itemize}

These scripts assume that 5 CUDA GPUs are available and run 5 benchmarks in parallel. If fewer GPUs are available, it is recommended to modify the scripts to run the benchmarking scripts sequentially (by removing the ampersand) and to adapt the \lstinline{--cuda=N} option.

%% file: appendix/unweighting.tex
Generating i.i.d points following an arbitrary probability distributions in high dimensions is not a priori a trivial task. A straightforward way to obtain such data is to use rejection sampling, which can be based on any distribution $q$ from which we can sample. Given an i.i.d sample $x_{1}, \dots, x_{N} \sim q$, we can define weights $w(x_{i}) = p(x_{i})/q(x_{i})$ and keep/reject points with probability $w(x_{i})/w_{\text{max}}$.

The main metric for evaluating the performance of this algorithm is the unweighting efficiency: how much data is kept from an original sample of size $N$ on average, which is expressed as

    \begin{equation}
      \label{eq:unweight_eff}
      \epsilon_\text{unw} = \underset{x\sim q}{\mathbb{E}} \frac{w(x)}{w_{\text{max}}}.
    \end{equation}


%% file: appendix/vegas_problems.tex
We define three functions over the two dimensional hypercube:

\begin{align}
    f_\text{camel}(x) &= \exp \left(-\left(\frac{x-\mu_1)}{\sigma}\right)^2\right) + \exp \left(-\left(\frac{x-\mu_2)}{\sigma}\right)^2\right), \label{eq:came_def}\\
    f_\varnothing(x) & =  
    \min \left[1,\ \exp \left( -\left( \frac{
    \left| x \right| - r
    }{\sigma_\varnothing}\right)^2  \right)
    +
    \exp \left( - \left(\frac{
        a\cdot x
    }{\sigma_\varnothing}\right)^2\right) \right]\\
    f_\text{sin}(x) & = \cos \left(k\cdot x \right)^2,
\end{align}

to which we refer respectively as the camel, slashed circle and sinusoidal target functions. We set their parameters as follows
\begin{equation}
    \mu_1 = \begin{pmatrix}
        0.25\\
        0.25
    \end{pmatrix}, \;
    \mu_2 = \begin{pmatrix}
        0.75\\
        0.75
    \end{pmatrix},\;
    a = \begin{pmatrix}
        1\\-1
    \end{pmatrix},\;
    k = \begin{pmatrix}
    6\\
    6    
    \end{pmatrix},\;
    \sigma=0.1,\;
    \sigma_\varnothing=0.1,\;
    r=0.3
\end{equation}
We chose these functions because they illustrate different failure modes of the VEGAS algorithm:
 
\begin{itemize}
    \item Because of the factorized PDF of VEGAS, the camel function leads to 'phantom peaks' in the off diagonal. This problem grows exponentially with the number of dimensions but can be addressed by a change of variable to align the peaks with an axis of integration.
    \item The sinuoidal function makes it nearly impossible for VEGAS to provide any improvement: the marginal PDF over each variable of integration is nearly constant. Again this type of issue can be addressed by a change of variable provided one knows how to perform it.
    \item The slashed circle function is an example of a hard-for-VEGAS function that cannot be improved significantly by a change of variables. One can instead use multi-channeling, but this requires a lot of prior knowledge and has a computational costs since each channel is its own integral.
\end{itemize}


%% file: performance/supp_figs.tex
\begin{figure}[!ht]
    \centering
    \includegraphics[width=\textwidth]{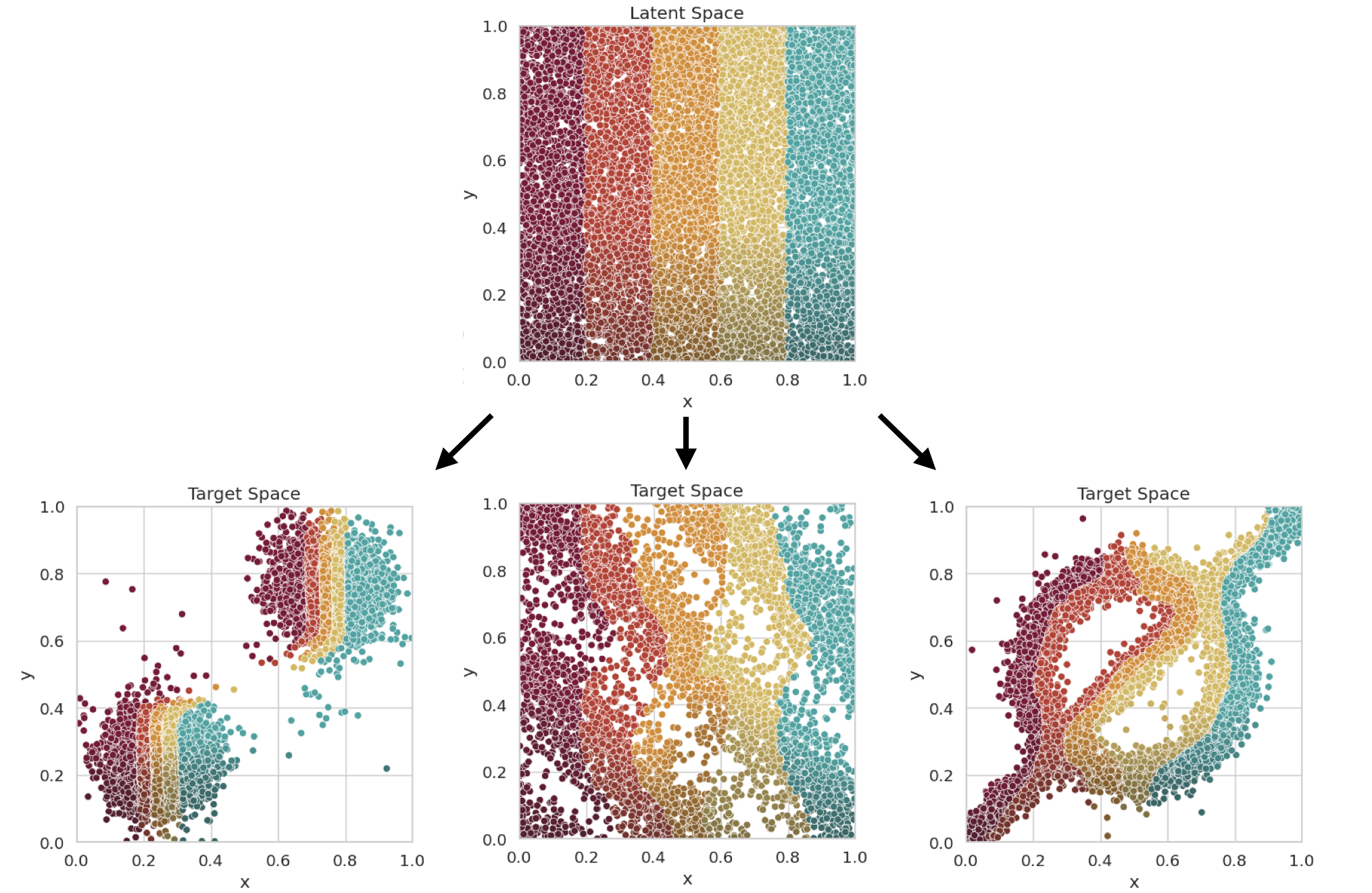}
    \caption{Mapping between the uniform point density in the latent space and the target distribution for the camel function, the sinusoidal function, the slashed circle function. Points are colored based on their position in latent space.}
    \label{fig:2dmapping}    
\end{figure}

%% file: appendix/benchmark_data.tex
\begin{table}[!ht]
    \centering
    \begin{tabular}{c|ccccc|cc}
        \hline
         & \multicolumn{5}{c|}{Dimension} & \multirow{2}{*}{$\sigma_\text{1d}$}&  \multirow{2}{*}{rel. st.d} \\
        & 2 & 4 & 8 & 16  & 32 & &\\\hline
        \parbox[t]{2mm}{\multirow{7}{*}{\rotatebox[origin=c]{90}{Camel param. $\sigma$}}}
         & $2.00 \times 10^{-2}$ &$8.93 \times 10^{-2}$ &$2.04 \times 10^{-1}$ &$3.61 \times 10^{-1}$ &$5.06 \times 10^{-1}$ & 0.001 & $14.1$\\
        & $6.32 \times 10^{-2}$ &$1.64 \times 10^{-1}$ &$3.14 \times 10^{-1}$ &$4.64 \times 10^{-1}$ &$6.06 \times 10^{-1}$ & 0.01 & $4.41$\\
        & $2.21 \times 10^{-1}$ &$3.78 \times 10^{-1}$ &$5.23 \times 10^{-1}$ &$6.67 \times 10^{-1}$ &$8.25 \times 10^{-1}$ & 0.1 & $1.22$\\
        & $4.51 \times 10^{-1}$ &$5.93 \times 10^{-1}$ &$7.43 \times 10^{-1}$ &$9.11 \times 10^{-1}$ &$1.10 $ & 0.3 & $0.56$\\
        & $6.44 \times 10^{-1}$ &$7.99 \times 10^{-1}$ &$9.74 \times 10^{-1}$ &$1.18 $ &$1.41 $ & 0.5 & $0.32$\\
        & $8.62 \times 10^{-1}$ &$1.05 $ &$1.26 $ &$1.51 $ &$1.81 $ & 0.7 & $0.19$\\
        & $1.21 $ &$1.45 $ &$1.73 $ &$2.07 $ &$2.47 $ & 1.0 & $0.10$
    \end{tabular}
    \caption{Setup of the 35 different camel functions considered to benchmark \zunis. We scan over function relative standard deviations, which correspond to different $\sigma$ parameters for each dimension(\cref{eq:came_def}). We provide the corresponding width of a 1D gaussian ($\sigma_{1d}$) with the same variance for reference.}
    \label{tab:camel_setups}
\end{table}

\begin{table}[!ht]
    \def\arraystretch{1.3}
    \setlength{\tabcolsep}{2pt}
    \centering
    \begin{tabular}{c|ccccc|cc}
        \hline
         & \multicolumn{5}{c|}{Dimension} & \multirow{2}{*}{$\sigma_\text{1d}$}&  \multirow{2}{*}{rel. st.d} \\
        & 2 & 4 & 8 & 16  & 32 & &\\\hline
        \parbox[t]{4mm}{\multirow{7}{*}{\rotatebox[origin=c]{90}{VEGAS speed-up}}}
        & $7.6^{+2.4}_{-2.4} \times 10^{1}$ & $6.0^{+1.2}_{-1.0} \times 10^{2}$ & $4.3^{+1.4}_{-1.0} \times 10^{3}$ & $9.0^{+4.7}_{-3.1} \times 10^{2}$ & $3.5^{+2.7}_{-1.8} \times 10^{0}$ & $0.001$  & $14.11$\\ 
        & $2.0^{+0.4}_{-0.4} \times 10^{2}$ & $5.8^{+0.8}_{-1.0} \times 10^{2}$ & $5.1^{+0.8}_{-0.9} \times 10^{2}$ & $2.8^{+0.3}_{-0.3} \times 10^{2}$ & $4.5^{+0.6}_{-0.8} \times 10^{1}$ & $0.01$  & $4.41$\\ 
        & $3.4^{+0.5}_{-0.3} \times 10^{2}$ & $1.0^{+0.1}_{-0.1} \times 10^{2}$ & $3.5^{+0.1}_{-0.2} \times 10^{1}$ & $1.9^{+0.1}_{-0.1} \times 10^{1}$ & $1.2^{+0.1}_{-0.1} \times 10^{1}$ & $0.1$  & $1.22$\\ 
        & $6.3^{+0.4}_{-0.3} \times 10^{1}$ & $2.0^{+0.0}_{-0.1} \times 10^{1}$ & $7.2^{+0.3}_{-0.3} \times 10^{0}$ & $2.8^{+0.1}_{-0.1} \times 10^{0}$ & $1.3^{+0.1}_{-0.0} \times 10^{0}$ & $0.3$  & $0.56$\\ 
        & $1.2^{+0.1}_{-0.1} \times 10^{1}$ & $3.5^{+0.1}_{-0.2} \times 10^{0}$ & $9.1^{+0.6}_{-0.4} \times 10^{-1}$ & $3.7^{+0.2}_{-0.2} \times 10^{-1}$ & $2.0^{+0.1}_{-0.1} \times 10^{-1}$ & $0.5$  & $0.33$\\ 
        & $2.4^{+0.2}_{-0.2} \times 10^{0}$ & $5.3^{+0.5}_{-0.4} \times 10^{-1}$ & $1.3^{+0.1}_{-0.1} \times 10^{-1}$ & $5.5^{+0.4}_{-0.3} \times 10^{-2}$ & $3.0^{+0.1}_{-0.2} \times 10^{-2}$ & $0.7$  & $0.20$\\ 
        & $4.0^{+0.6}_{-0.5} \times 10^{-1}$ & $7.8^{+0.6}_{-0.5} \times 10^{-2}$ & $1.6^{+0.1}_{-0.1} \times 10^{-2}$ & $7.8^{+0.8}_{-0.6} \times 10^{-3}$ & $4.8^{+0.3}_{-0.4} \times 10^{-3}$ & $1.0$  & $0.11$\\        
    \end{tabular}
    \caption{Variance reduction factor compared to VEGAS for each of the 35 different camel setups defined in \cref{tab:camel_setups}.}
    \label{tab:camel_default_results}
\end{table}


%% file: appendix/flat_matrix_element.tex
\begin{figure}[!ht]
    \centering
    \begin{subfigure}[b]{0.49\textwidth}
        \centering
        \includegraphics[width=\textwidth]{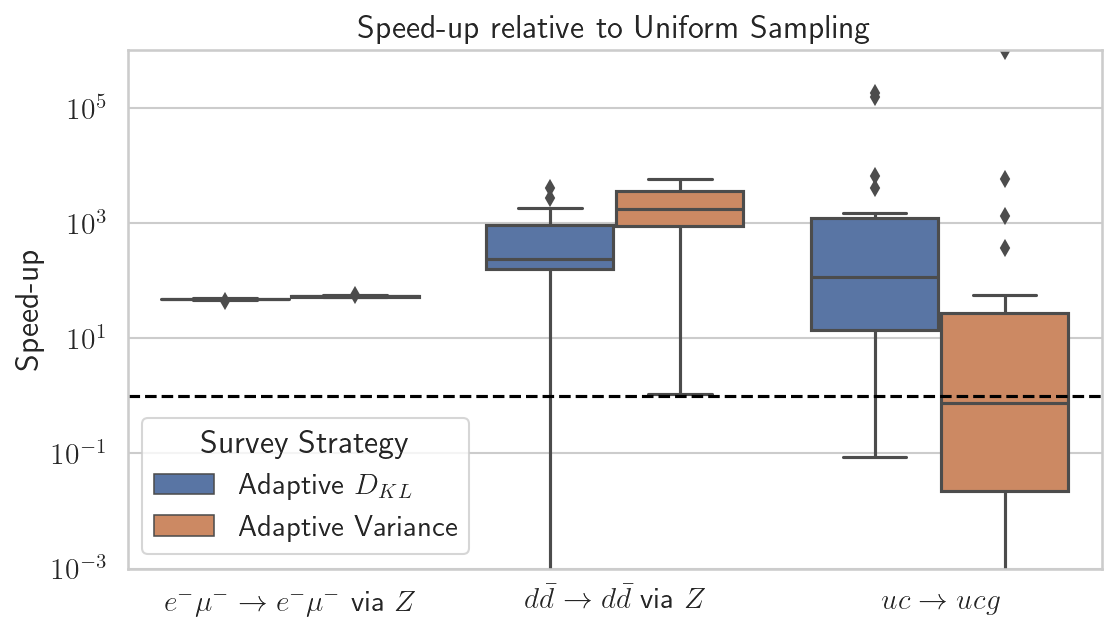}
        \caption{}
        \label{fig:madgraph_comparison_speed_flat}
    \end{subfigure}
    \begin{subfigure}[b]{0.49\textwidth}
        \centering
        \includegraphics[width=\textwidth]{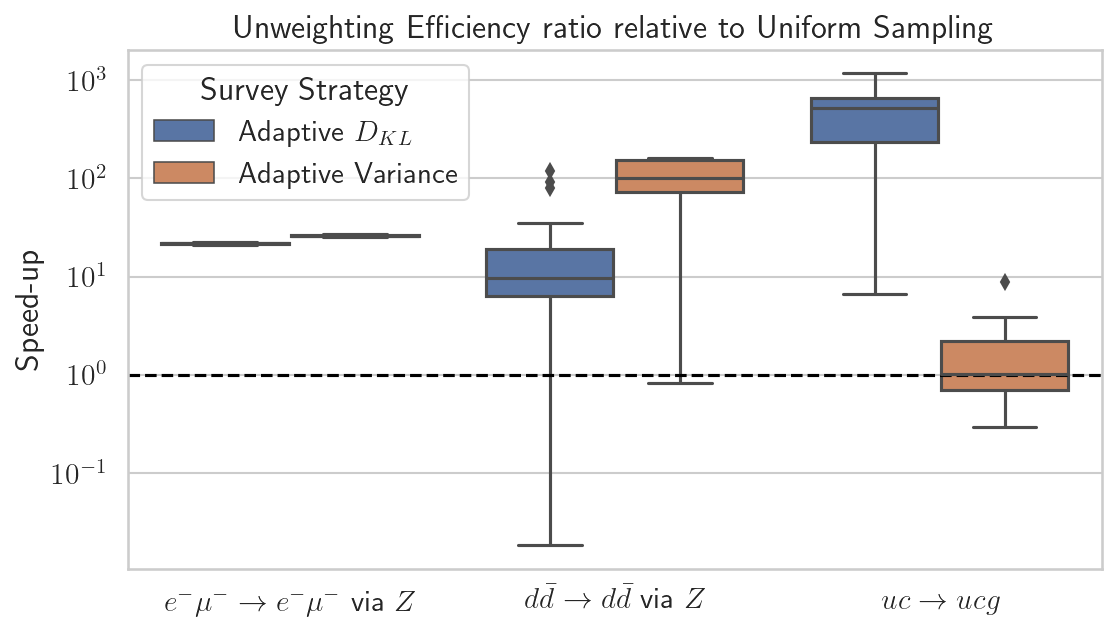}
        \caption{}
        \label{fig:madgraph_comparison_uwr_flat}
    \end{subfigure}
    \caption{Average performance of \zunis over 20 runs relative to flat sampling, measured by the relative speed-up in \ref{fig:madgraph_comparison_speed_flat} and by the relative unweighting efficiency in \ref{fig:madgraph_comparison_uwr_flat}.}
    \label{fig:madgraph_comparison_flat}
\end{figure}

%% file: appendix/survey_strategies.tex
In the following, we want to investigate how the integration of the three example processes in \ref{sec:madgraph} with \zunis behaves relative to VEGAS in dependence of the choice of the loss function, the survey strategy and the number of epochs during training. For all other options, the default values are chosen again, except for the number of points during the survey phase, which is set to 500,000.

\begin{figure}[!ht]
    \centering
    \begin{subfigure}[b]{\textwidth}
        \centering
        \includegraphics[width=\textwidth]{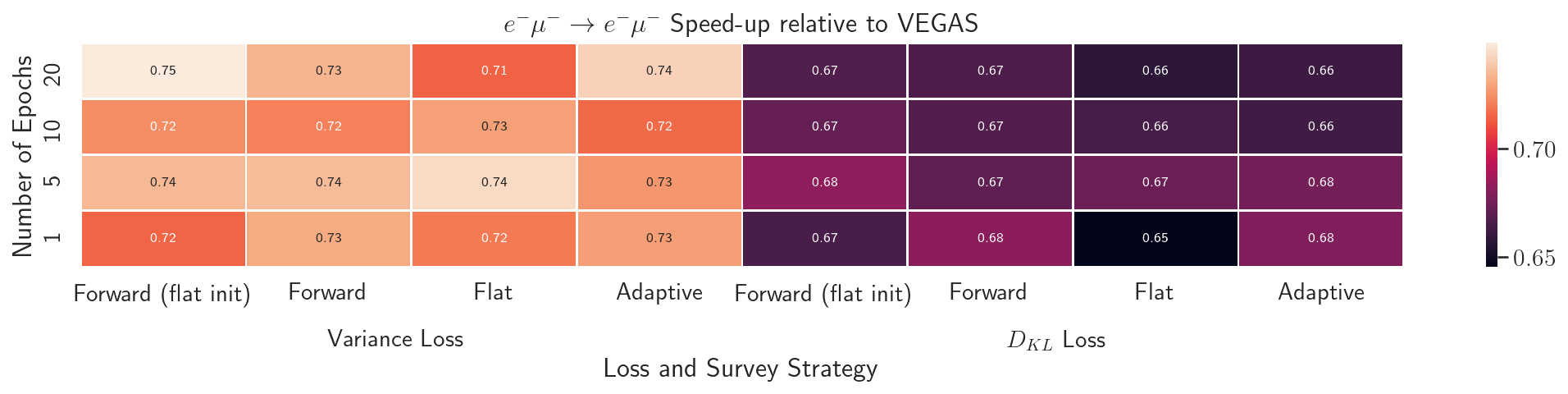}
        \caption{}
        \label{fig:emu_var}
    \end{subfigure}
    \begin{subfigure}[b]{\textwidth}
        \centering
        \includegraphics[width=\textwidth]{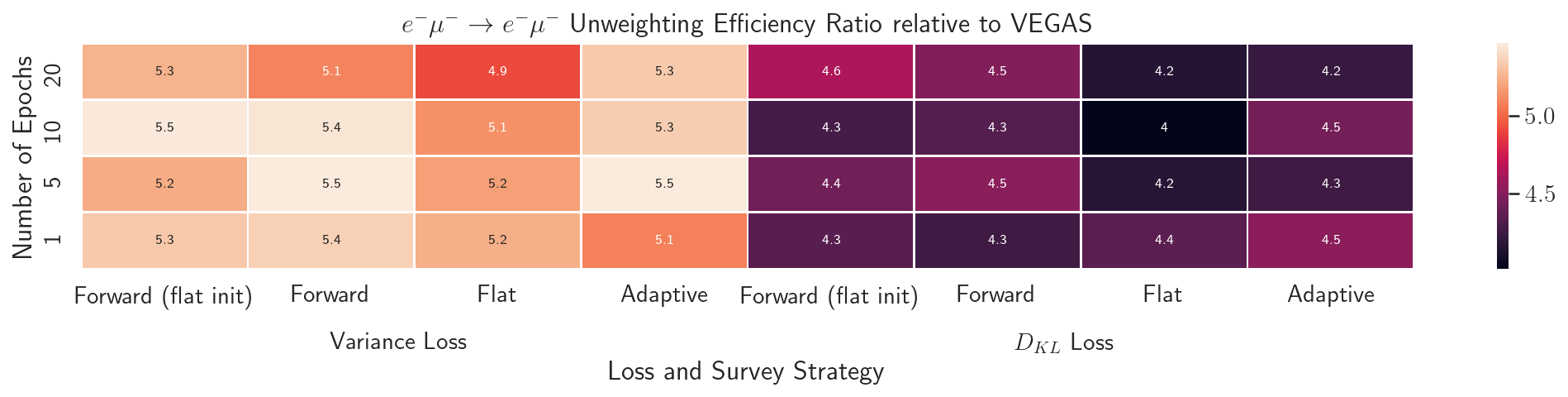}
        \caption{}
        \label{fig:emu_uwr}
    \end{subfigure}
    \caption{Median of the performance of \zunis over 20 runs relative to VEGAS for the process $ e^- \mu \rightarrow e^- \mu$ via $Z$  depending on the loss function, the survey strategy and number of epochs, measured by the relative speed-up in \ref{fig:emu_var} and by the relative unweighting efficiency in \ref{fig:emu_uwr}.}
    \label{fig:emu}
\end{figure}

Figure \ref{fig:emu_var} shows that for a simple process like $ e^- \mu \rightarrow e^- \mu$ via $Z$ , where no correlations exist, \zunis cannot reach the speed-up achieved by VEGAS. Variance loss seems to lead to higher variance improvements than $D_{KL}$ loss. Contrary, \ref{fig:emu_uwr} shows that \zunis can greatly improve the unweighting efficiency for this process. The effect is again consistently stronger when using variance loss. Using a flat survey strategy suffers for both loss functions from overfitting, whereas adaptive sampling in average performs slightly better and does not show overfitting.

$ d \bar d \rightarrow d \bar d$ via $Z$ presents a more realistic use case, as the parton distribution functions introduce correlations between the integration variables which present a challenge to the VEGAS algorithm. 

\begin{figure}[!ht]
    \centering
    \begin{subfigure}[b]{\textwidth}
        \centering
        \includegraphics[width=\textwidth]{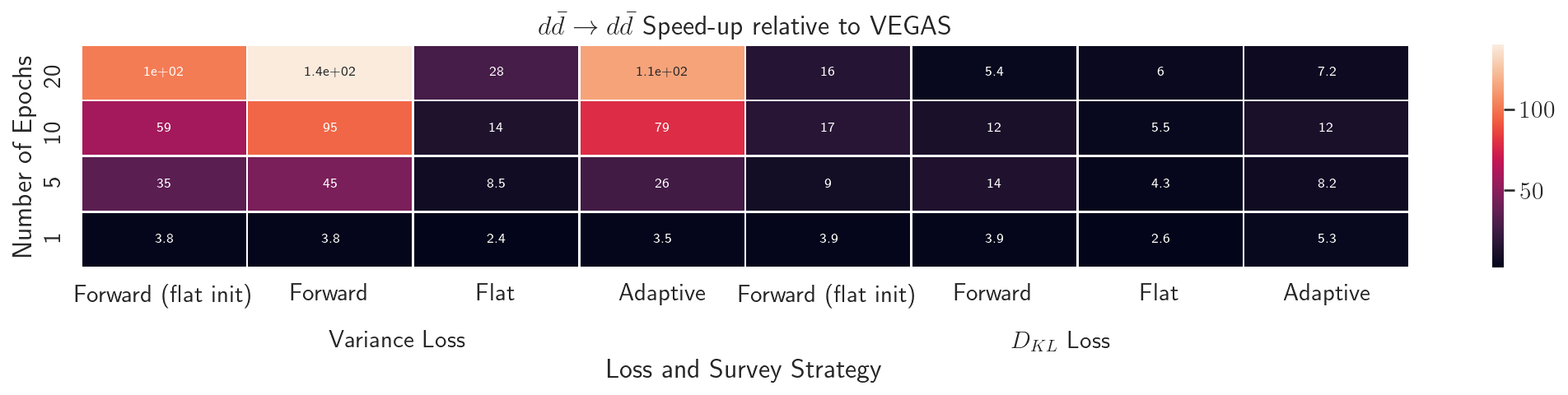}
        \caption{}
        \label{fig:d_var}
    \end{subfigure}
    \begin{subfigure}[b]{\textwidth}
        \centering
        \includegraphics[width=\textwidth]{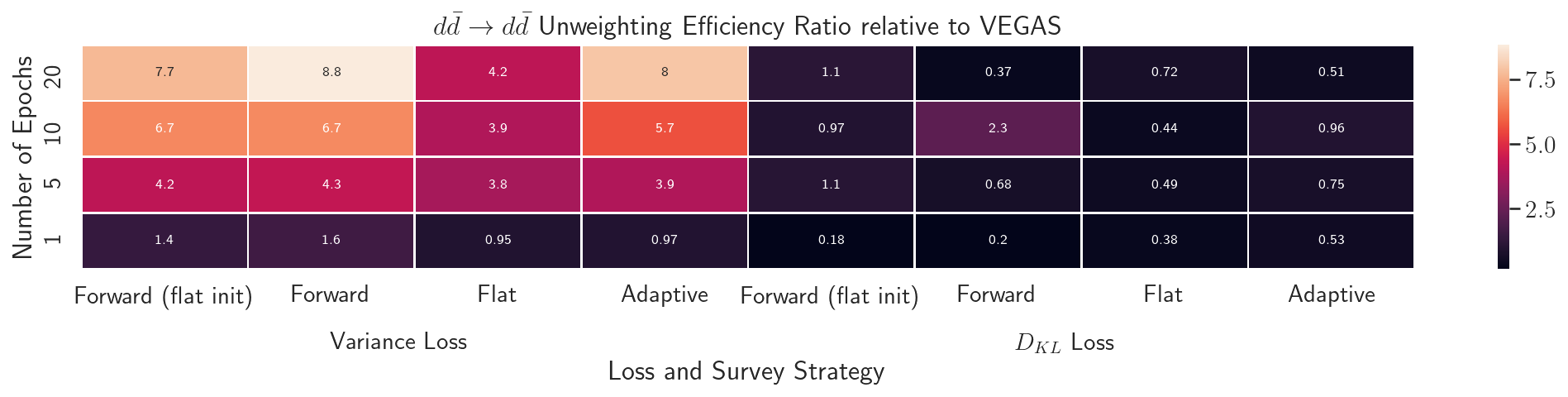}
        \caption{}
        \label{fig:d_uwr}
    \end{subfigure}
    \caption{Median of the performance of \zunis over 20 runs relative to VEGAS for the process $ d \bar d \rightarrow d \bar d$ via $Z$  depending on the loss function, the survey strategy and number of epochs, measured by the relative speed-up in \ref{fig:d_var} and by the relative unweighting efficiency in \ref{fig:d_uwr}.}
    \label{fig:d}
\end{figure}

For this process, both the speed-up and the unweighting efficiency ratio clearly favor the variance loss again, which outperforms in both metrics the $D_{KL}$ loss when multiple epochs are used, as can be seen in \ref{fig:d}. The richer structure of the integrand reduces the effect of overfitting. Therefore, the performance increases or stays approximately constant except for the combination of $D_{KL}$ loss and the forward survey strategy. For the variance loss, it becomes apparent that the flat survey strategy increases much slower in performance than alternative strategies as a function of the number of epochs. However, for combination of losses and survey strategies, VEGAS could be outperformed substantially in terms of integration speed.

\begin{figure}[!ht]
    \centering
    \begin{subfigure}[b]{\textwidth}
        \centering
        \includegraphics[width=\textwidth]{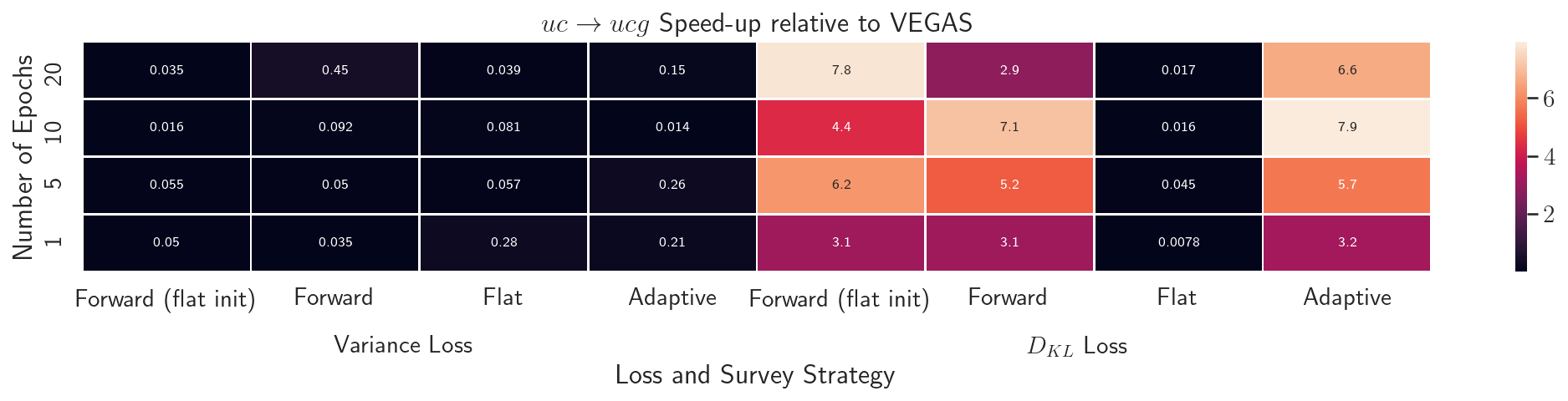}
        \caption{}
        \label{fig:u_var}
    \end{subfigure}
    \begin{subfigure}[b]{\textwidth}
        \centering
        \includegraphics[width=\textwidth]{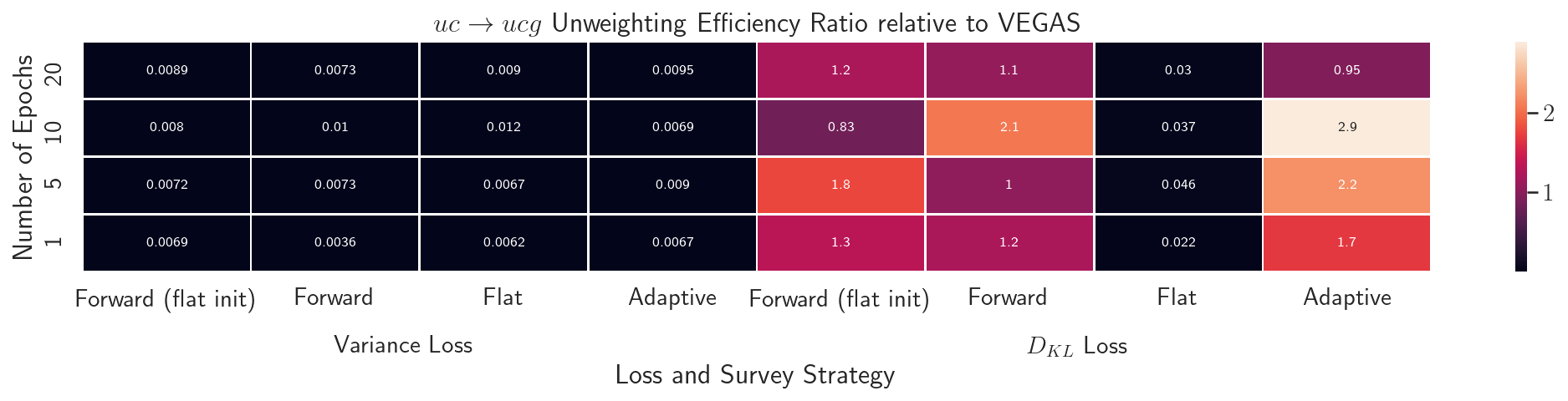}
        \caption{}
        \label{fig:u_uwr}
    \end{subfigure}
    \caption{Median of the performance of \zunis over  20 runs relative to VEGAS for the process $ u c \rightarrow u c g$  depending on the loss function, the survey strategy and number of epochs, measured by the relative speed-up in \ref{fig:u_var} and by the relative unweighting efficiency in \ref{fig:u_uwr}.}
    \label{fig:u}
\end{figure}

An opposite picture is drawn by the process $ u c \rightarrow u c g$ in figure \ref{fig:u}, for which, apart from the flat survey strategy, $D_{KL}$ loss is in general favored both for integration speed as well as unweighting efficiency ratio. The adaptive survey strategy is here giving the best results, although for a high number of epochs causes overfitting for the unweighting efficiency.

The take-home message of this section is, one the one hand, that the flat survey strategy is in general not recommended. Apart from this, the most important mean to improve the quality of importance sampling are testing whether, independent of the survey strategy, the loss function should be chosen differently.

%% file: appendix/optimal_is.tex
Let us show that the optimal probability distribution for importance sampling is the function itself. 
Explicitly, as discussed in \cref{sec:importance_sampling}, we want to find the probability distribution $p$ defined over some domain $\Omega$ which minimizes the variance of the estimator $f(X)/p(X)$, for $X\sim p(X)$. We showed that this amounts to solving the following optimization problem:
\begin{equation}
\begin{split}
    \underset{p}{\min}\,\mathcal{L}(p) = \int_\Omega dx \frac{f(x)^2}{p(x)}, \quad
    \text{ such that } \int dx \, p(x) = 1,
\end{split}
\end{equation}
which we can encode using Lagrange multipliers
\begin{equation}
   p = \arg\min \mathcal{L}(p,\lambda) = \int_\Omega dx \frac{f(x)^2}{p(x)} + \lambda \left(p(x) - \frac{1}{V(\Omega)}\right).
\end{equation}
We can now solve this problem by finding extrema with functional derivatives
\begin{equation}
    \frac{\delta \mathcal{L}(p,\lambda) }{\delta p(x)} = \lambda - \frac{f(x)^2}{p(x)^2},
\end{equation}
which indeed is zero if $p(x) \propto \left| f(x) \right|$. Furthermore, this extremum is certainly a minimum because the loss function is positive and unbounded. Indeed, if we separate $\Omega$ into two disjoint measurable subdomains $\Omega_1$ and $\Omega_2$, and define $p_\alpha(x)$ such that points are drawn uniformly over $\Omega_1$ with probability $\alpha$ and uniformly over $\Omega_2$ with probability $1-\alpha$, then the resulting loss function would be
\begin{equation}
    \mathcal{L}(p_\alpha) = \frac{V(\Omega_1)}{\alpha} \int_{\Omega_1} dx\, f(x)^2 + \frac{V(\Omega_2)}{1-\alpha}\int_{\Omega_2} dx\, f(x)^2,
\end{equation} 
which can be made arbitrarily large by sending $\alpha$ to 0.


%% file: framework/flows.tex
\paragraph{Flows map batches of points and their densities.}
The \zunis library implements normalizing flows by specifying a general interface defined as a Python abstract class: \texttt{GeneralFlow}.
All flow models in \zunis are child classes of \texttt{GeneralFlow}, which itself inherits from the Pytorch \texttt{nn.Module} interface.

As defined in \cref{sec:normalizing_flows}, a normalizing flow in \zunis is a bijective mapping between two $d$ dimensional spaces, which in practice are always the unit hypercube $[0,1]^d$ or $\mathbb{R}^d$, with a tractable Jacobian so that it can be used to map probability distributions.
To this end, the \texttt{GeneralFlow} interface defines normalizing flows as a callable Python object which acts on batches of point drawn from a known PDF $p$. A batch of points $x_i$ with their PDF values is encoded as a Pytorch \texttt{Tensor} object $X$ organized as follows
\begin{align}
    X &= (X_1,\dots, X_\text{batch})\in \mathbb{R}^\text{batch}\times \mathbb{R}^{d+1},
\end{align}
where each $X_i$ corresponds to a points stacked with its negative log density
\begin{equation}
    X_i = \begin{pmatrix}
        x_{i,1}\\
        \vdots\\
        x_{i,d}\\
        -\log p(x_i)
    \end{pmatrix}.
\end{equation}
Encoding point densities by their negative logarithm makes their transformation under normalizing flows additive. Indeed if we have a mapping $Q$ with Jacobian determinant $j_Q$, then $x\sim p(x)$ is mapped to $y=Q(x)\sim \tilde p(y)$ such that
\begin{equation}
    - \log \tilde p(y) = - \log p(x) + \log j_Q(x).
\end{equation}

\paragraph{Coupling Cells are flows defined by an element-wise transform and a mask.}

All flow models used in \zunis in practice are implemented as a sequence of coupling cell transformations acting on a subset of the variables. The abstract class \texttt{GeneralCouplingCell} and its child \texttt{InvertibleCouplingCell} specifies the general organization of coupling cells as needing to be instantiated with

\begin{itemize}
    \item a dimension $d$
    \item a mask defined as a list of boolean specifying which coordinates are transformed or not
    \item a transform that implements the mapping of the non-masked variables
\end{itemize}

In release v1.0 of \zunis two such classes are provided: \texttt{PWLinearCoupling} and \texttt{PWQuadraticCoupling}, which respectively implement the piecewise linear and piecewise quadratic coupling cells proposed in \citep{mller2018neural}.
New coupling cells can easily be implemented, as explained in \cref{app:new_cells}.
Both existing classes rely on dense neural networks for the prediction of the shape of their one-dimensional piecewise-polynomial mappings, whose parameters are set at instantiation. 

Here is how one can use a piecewise-linear coupling cell for sampling points

\begin{lstlisting}
import torch
from zunis.models.flows.coupling_cells.piecewise_coupling.piecewise_quadratic import PWQuadraticCoupling
d=2
N_batch=10
mask = [True,False]
x = torch.zeros((N_batch,d+1))
# Sample the d first entries uniformly, keep 0. for the negative log jacobian
x[...,:-1].uniform_()
print(x[0]) # [0.3377, 0.4362, 0.]
f = PWQuadraticCoupling(d=d,mask=mask)
y = f(x)
print(y[0]) # [0.3377, 0.4411, -0.0314]
\end{lstlisting}

We provide further details of the use and possible parameters of flows in the documentation of \zunis: \url{ https://zunis.readthedocs.io/en/stable/}.


%% file: framework/trainers.tex
The design of the \zunis library intentionally restricts \texttt{Flow} models to being bijective mappings instead of being ways to evaluate and sample from PDFs so as not to restrict their applicability (see \cite{brehmer2020flows} for an example).
The specific application in which one uses a normalizing flow, and in our case how precisely one samples from it, is intimately linked to how such a model is trained. 
As a result, \zunis bundles together the low-level training tools for \texttt{Flow} models together with sampling tools inside the \texttt{Trainer} classes.

The general blueprint for such classes is defined in the \texttt{GenericTrainerAPI} abstract class while the main implementation for users is provided as \texttt{StatefulTrainer}. At instantiation, all trainers expect a \texttt{Flow} model and \texttt{flow\_prior} which samples point from a fixed PDF in latent space.
These two elements together define a probability distribution over the target space from which one can sample.

There are two main ways one interacts with \texttt{Trainers}:
\begin{itemize}
    \item One can sample points from the PDF defined by the model and the prior using the \texttt{sample\_forward} method.
    \item One can train over a pre-sample batch of points, their sampling PDF and the corresponding function values using \texttt{train\_on\_batch(self, x, px, fx)}
\end{itemize}

In practice, we expect that the main way users will use \texttt{Trainers} is for sampling pre-trained models. In the context of particle physics simulations for example, \textit{unweighting} is a common task, which aims at sampling exactly from a known function $f$. A common approach is the \textit{Hit-or-miss} algorithm~\citep{James:1980yn}, whose efficiency is improved by sampling from a PDF approaching $f$. This is how one would use a trainer trained on $f$:

\begin{lstlisting}
# [...]
# import or train a trainer `pretrained_trainer` 
import torch

# Sampling points
xj = pretrained_trainer.sample_forward(100)
x = x[:, :-1]
px = (-x[:,-1]).exp()
fx = f(x)

# Applying the veto algorithm
fmax = fx.max()
veto = (fx/fmax - torch.zeros_like(fx).uniform_(0.,1.)) > 0.
x_unweighted = x[veto]
# x_unweighted follows the PDF obtained by normalizing f.
\end{lstlisting}

%% file: framework/integrators.tex
Integrators are intended as the main way for standard users to interact with \zunis. They provide a high-level interface to the functionalities of the library and only optionally require users to know to what lower levels of abstractions really entail and to what their options correspond.
From a practical point of view, the main interface of \zunis for integration is implemented as the \texttt{Integrator}, which is a factory function that instantiates the appropriate integrator class based on a choice of options.

All integrators follow the same pattern, defined in the \texttt{SurveyRefineIntegratorAPI} and \texttt{BaseIntegrator} abstract classes. Integration start by performing a survey phase, in which it optimizes the way it samples points and then a refine phase, in which it computes the integral by using its learned sampler. Each phase proceeds through a number of steps, which can be set at instantiation or when integrating:

\begin{lstlisting}
# Setting values at instantiation time
integrator = Integrator(d=d, f=f, n_iter_survey=3, n_iter_refine=5)
# Override at integration time
integral_data = integrator.integrate(n_survey=10, n_refine=10)
\end{lstlisting}

For both the survey and the refine phases, using multiple steps is useful to monitor the stability of the training and of the integration process: if one step is not within a few standard deviations of the next, either the sampling statistics are too low, or something is wrong. For the refine stage, this is the main real advantage of using multiple steps. On the other hand, at each new survey step, a new batch of points is re-sampled, which can be useful to mitigate overfitting.

By default, only the integral estimates obtained during the refine stage are combined to compute the final integral estimate, and their combination is performed by taking their average. Indeed, because the model is trained during the survey step, the points sampled during the refine stage are correlated in an uncontrolled way with the points used during training. Ignoring the survey stage makes all estimates used in the combination independent random variables, which permits us to build a formally correct estimator of the variance of the final result.

%% file: appendix/new_couplings.tex
To implement a new invertible coupling cell inheriting from \texttt{InvertibleCouplingCell}, one must provide an \texttt{InvertibleTransform} object and define a callable attribute \texttt{T} computing the parameters of the transform. For example, consider a very simple linear coupling cell over $\mathbb{R^d}$
\begin{align}
    y = Q(x):\; \left\{ 
        \begin{array}{l}
            y^A = x^A\\
            y^B = \exp\left(T(x^A)\right) \times x^B,
        \end{array}
     \right. 
\end{align}
where $T(x^A)$ is a scalar strictly positive value. This can be defined in the following way in \zunis

\begin{lstlisting}
import torch
from zunis.models.flows.coupling_cells.general_coupling import InvertibleCouplingCell
from zunis.models.flows.coupling_cells.transforms import InvertibleTransform
from zunis.models.layers.trainable import ArbitraryShapeRectangularDNN

class LinearTransform(InvertibleTransform):
    def forward(self,x,T):
        alpha = torch.exp(T)
        logj = T*x.shape[-1] 
        return x*alpha, logj
    def backward(self,x,T):
        alpha = torch.exp(-T)
        logj = -T*x.shape[-1] 
        return x*alpha, logj
\end{lstlisting}

\begin{lstlisting}
    
class LinearCouplingCell(InvertibleCouplingCell):
    def __init__(self, d, mask, nn_width, nn_depth):
        transform = LinearTransform()
        super(LinearCouplingCell, self).__init__(d=d, mask=mask,transform=transform)
        d_in = sum(mask)
        self.T = ArbitraryShapeRectangularDNN(d_in=d_in,
                                              out_shape=(1,),
                                              d_hidden=nn_width,
                                              n_hidden=nn_depth)
\end{lstlisting}


%% file: appendix/hardware.tex
The computations presented in this work were performed on a computing cluster using a Intel(R) Xeon(R) Gold 5218 CPU @ 2.30GHz with 376 GB RAM. Processes which could be performed on the GPU were done on a GeForce RTX 2080 having 12 GB memory and running on CUDA 11.0.